\documentclass{article}
\usepackage{amsmath}
\usepackage{amsfonts}
\usepackage{fontsize}
\changefontsize[14pt]{12pt}
\usepackage{setspace} 
\usepackage{geometry} 
\usepackage{booktabs}
\usepackage{rotating}
\usepackage{calc} 
\usepackage{array}
\usepackage{graphicx} 
\usepackage{caption}
\usepackage{subcaption}
\graphicspath{{./images/}}
\usepackage{tabularx}
\usepackage{ragged2e} 
\usepackage{float}
\usepackage{caption}

\usepackage{authblk}

\title{Null Geodesics in the Reissner-Nordstr\"{o}m Black Hole Spacetime Pierced by a Cosmic String}
\author{Tianxu Huo}

\affil{College of Physics, Nanjing University of Aeronautics and Astronautics, Nanjing, 210016, China}

\date{}

\begin{document}
	
	\maketitle
	\begingroup
	\renewcommand{\thefootnote}{}
	\footnotetext{E-mail: 23010702115@usx.edu.cn}
	\endgroup
	\setcounter{footnote}{0}   
	
     \begin{quote}
		\noindent\textbf{Abstract:} In this paper, we investigate null geodesics in the spacetime of a Reissner-Nordström black hole pierced by a cosmic string. Analysis of the equations of motion reveals that null geodesics in this conical spacetime can be classified into three main categories, with the classification entirely determined by the ratio $|Q|/M$ (where $Q$ and $M$ denote the black hole charge and mass, respectively). Analytical solutions to the null geodesic equations are obtained for each category. Based on these solutions, we numerically simulate light trajectories and analyze in detail the effects of the black hole charge $Q$ and the cosmic string parameter $\alpha$. It is found that the cosmic string parameter $\alpha$ enhances the winding behavior of null geodesics across all three main categories. Finally, we consider light deflection in the equatorial plane and derive an expression for the deflection angle, revealing that the charge $Q$ decreases the deflection angle, whereas the cosmic string parameter $\alpha$ increases it.

	\end{quote}
	
	\newpage

	\section*{1. Introduction}
	
	The concept of cosmic strings was first proposed by Kibble in 1976 \cite{1}, originating from the spontaneous symmetry breaking mechanism in grand unified gauge theories of particle physics. According to this theory, during the evolution of the early universe, when the vacuum undergoes spontaneous symmetry breaking, it behaves like a typical phase transition system, retaining certain remnants of the original symmetry in the previously symmetric false vacuum, thereby forming topological defects \cite{2}. When the symmetry breaking results in an infinitely long one-dimensional linear structure, such a topological defect is called a cosmic string \cite{3}. Cosmic strings can exist as infinitely long straight strings or closed loops, forming a network in the universe that consists of both infinite strings and closed loops. Studies suggest that cosmic strings may be related to the formation of large-scale structure in the universe \cite{4,5}. Moreover, cosmic strings can produce several observable effects. Efforts to detect cosmic strings have utilized various observable phenomena, in particular the development of cusps and kinks on strings, which can radiate strong bursts of gravitational waves \cite{6,7}. The distinctive gravitational wave signatures of cosmic strings have attracted considerable theoretical and observational interest, and several studies have set upper limits on the cosmic string tension using gravitational wave observations (see details in \cite{8,9,10}). In addition to gravitational waves, the microlensing effect caused by cosmic strings is also considered a promising detection approach \cite{11}.
	
	Cosmic strings may not exist in isolation. Considering the epoch of primordial black hole formation, regions of high density within a cosmic string network could become sufficiently dense to allow one or more cosmic strings to attach to a collapsing region that is about to form a primordial black hole \cite{12}. Another possible string model consists of a flux tube of a confined gauge field \cite{13}. Above the transition temperature, a black hole with a nonzero magnetic charge possesses a spherically symmetric Coulomb-type field. If the system is then cooled below the transition temperature, the field becomes confined, and strings emanating from the black hole are formed. Gauss's theorem requires that the total flux carried by the strings must equal the net flux across the black hole's horizon before the transition. Numerous properties of black holes pierced by cosmic strings have been reported, including black hole thermodynamics \cite{14,15,16}, geodesics \cite{17,18,19,20,21}, quasinormal modes \cite{22}, and strong-field gravitational lensing effects \cite{23,24}. Interestingly, \cite{25} show that a rigidly rotating string can extract the rotational energy from a rotating black hole, distinct from the conventional Penrose process, which extracts energy through particle decay within the ergosphere. Another intriguing study \cite{16} examined the influence of cosmic strings on the maximal interior volume and the entropy of the interior scalar field in quantum-corrected Schwarzschild black holes. The results show that, compared to the case of a pure quantum-corrected black hole, the presence of a cosmic string not only alters the black hole's interior entropy but also modifies the evolutionary relationship between the interior entropy and the Bekenstein-Hawking entropy for this topological-defect black hole. Nevertheless, during Hawking radiation, the total variation of these two types of entropy always satisfies the second law of thermodynamics---regardless of whether the quantum-corrected black hole is pierced by a cosmic string or not. A recent noteworthy study \cite{26}, for the first time, analyzed gravitational-wave strain data using waveforms constructed from numerical simulations of cosmic string loops collapsing to Schwarzschild black holes under strong gravity, and validated their approach using GW190521 as an example. The authors found that if only the ringdown signal is observed, a collapsing cosmic string loop can mimic a high-mass binary black hole merger.
	
	The geodesic motion of test particles provides an excellent probe of the curvature effects of spacetime. It is well known that some of the earliest and most celebrated confirmations of general relativity were based on geodesics, such as light deflection and the perihelion precession of Mercury. For a review of geodesic solutions in various classical spacetimes, see Ref.~\cite{27} and the references therein. In fact, the analytical solutions of the geodesic equations in the Reissner-Nordstr\"{o}m (RN) black hole spacetime were first given by Gackstatter \cite{28} as late as 1983. The geodesic motion in a Schwarzschild spacetime pierced by a cosmic string was first analyzed in Ref.~\cite{17}. Subsequently, Gal'tsov and Masar \cite{18} conducted a detailed study of geodesics in flat conical spacetime, conical Schwarzschild spacetime, and conical Lense--Thirring spacetime. They pointed out that, although the spacetime is only globally axisymmetric, locally there exist three Killing vectors forming an $SO(3)$ algebra, and on this basis analyzed the geodesic behavior in flat conical spacetime. Furthermore, they investigated geodesics in the conical Schwarzschild spacetime, discovering that the angular momentum vector of non-equatorial orbits precesses uniformly around the cosmic string axis, and also analyzed small oscillations of the orbits. Nevertheless, \cite{18} did not provide a systematic study of all possible geodesics, and the solutions to the geodesic equations involve elliptic integrals, which were not addressed in \cite{18}. In view of this, Ref.~\cite{20} provided a systematic classification of all possible types of geodesics in conical Schwarzschild spacetimes. However, for the more general case of a conical RN black hole spacetime, which significantly enriches the physics of black holes due to the presence of charge, no such study has been reported to date. The present work aims to fill this gap. We focus on null geodesics, provide a systematic classification of their possible types, derive analytical solutions for each class, and illustrate their properties using numerical methods.
	
	The paper is organized as follows. In Sec.2, we introduce the RN black hole spacetime pierced by a cosmic string. Sec.3 is devoted to the classification of null geodesics based on the equations of motion, followed by the derivation of analytical solutions in Sec.4. In Sec.5, we employ numerical methods to visualize the trajectories and analyze their physical properties. Finally, conclusions and outlooks are presented in Sec.~6.

	\section*{2. RN black holes pierced by a cosmic string}
	
	For an RN black hole pierced by a straight cosmic string, the line element is obtained by incorporating the deficit angle effect into the standard metric, and takes the form [2,29]:
	\begin{equation}
		ds^2 = -\left(1 - \frac{2M}{r} + \frac{Q^2}{r^2}\right) dt^2 + \left(1 - \frac{2M}{r} + \frac{Q^2}{r^2}\right)^{-1} dr^2 + r^2 \left(d\theta^2 + \alpha^2 \sin^2\theta d\phi^2\right).\tag{2.1}
	\end{equation}
	Here, $Q$ and $M$ denote parameters related to the electric charge and mass of the black hole, respectively, and $\alpha = 1 - 4\mu$, where $\mu$ is the linear mass density of the cosmic string. The deficit parameter $\alpha$ reduces the azimuthal angle range from $2\pi$ to $2\pi\alpha$, thereby endowing the spacetime with a conical geometry. Electromagnetic potential $A_\mu$ of this black hole solution is identical to that of the RN black hole [29]: $A_\mu = (Q/r, 0, 0, 0)$. Setting $Q^2 = \varepsilon M^2$ for convenience in the subsequent geodesic calculations leads to the following form of the metric
	\begin{equation}
		ds^2 = -\left(1 - \frac{2M}{r} + \frac{\varepsilon M^2}{r^2}\right) dt^2 + \left(1 - \frac{2M}{r} + \frac{\varepsilon M^2}{r^2}\right)^{-1} dr^2 + r^2 \left(d\theta^2 + \alpha^2 \sin^2\theta d\phi^2\right).\tag{2.2}
	\end{equation}
	In natural units, both $Q$ and $M$ have dimensions of length. With the reparameterization $Q^2 = \varepsilon M^2$, the dimensionless parameter $\varepsilon$ quantifies the relative strength of the electric charge, ranging over $(0, 1]$. The extremal case corresponds to $\varepsilon = 1$. From the metric (2.2), the inner and outer horizon radii are readily obtained as $r_{\pm} = M(1 \pm \sqrt{1-\varepsilon})$. It is evident that the presence of the cosmic string does not alter the locations of the horizons. However, due to the angular deficit induced by the cosmic string, their respective horizon areas are modified to
	\begin{equation}
		A_{\pm} = \int_{0}^{\pi} r_{\pm}^2 \sin\theta d\theta \int_{0}^{2\pi} \alpha d\phi = 4\pi \alpha r_{\pm}^2.\tag{2.3}
	\end{equation}
	Likewise, the equatorial proper lengths of the inner and outer horizons are reduced by the cosmic string: $l_{\pm} = \int_{0}^{2\pi} \sqrt{g_{\phi\phi}} d\phi = 2\pi \alpha r_{\pm}$. It is worth noting that, due to the angular deficit, the parameter $M$ should not be directly identified with the black hole’s physical mass $M_\text{phy}$. To demonstrate this explicitly, we evaluate the Komar mass, defined as
	\begin{equation}
		E_K = \frac{1}{4\pi} \int_{\partial\Sigma} d^2x \sqrt{h^{(2)}} n_{\mu} \sigma_{\nu} \nabla^{\mu} \xi_{(t)}^{\nu}.\tag{2.4}
	\end{equation}
	Here, $\xi_{(t)} = \partial/\partial t$ is the timelike Killing vector, $\partial\Sigma$ denotes the two-dimensional surface at infinity, and $n_{\mu}$ and $\sigma_{\nu}$ are the timelike and spacelike unit normal vectors to this surface, with non-vanishing components given respectively by
	\begin{equation}
		n_0 = -\left(1 - \frac{2M}{r} + \frac{\varepsilon M^2}{r^2}\right)^{1/2}, \quad \sigma_1 = \left(1 - \frac{2M}{r} + \frac{\varepsilon M^2}{r^2}\right)^{-1/2}.\tag{2.5}	
	\end{equation}
	From the black hole metric, we have
	\begin{equation}
		n_{\mu} \sigma_{\nu} \nabla^{\mu} \xi_{{(t)}}^{\nu} = n_0 \sigma_1 \nabla^0 \xi_{(t)}^1 = \frac{M}{r^2} - \frac{\varepsilon M^2}{r^3}.\tag{2.6}
	\end{equation}
     Consequently, the Komar mass contained within the radius $r$ is found to be

	\begin{equation}
		E_K(r) = \frac{1}{4\pi} \int_{0}^{\pi} r^2 \left(\frac{M}{r^2} - \frac{\varepsilon M^2}{r^3}\right) \sin\theta d\theta \int_{0}^{2\pi} \alpha d\phi = \alpha M - \frac{\alpha \varepsilon M^2}{r}.\tag{2.7}
	\end{equation}
	As $r \to \infty$, we have $\lim_{r \to \infty} E_K(r) = \alpha M$, indicating that the physical mass should be $M_\text{phy} = \alpha M$.
	
	Next, by constructing the embedding diagram of the equatorial plane ($\theta = \pi/2$), we examine the geometric  effect of the cosmic string. Setting $t = \text{const}$, the line element reduces to
	\begin{equation}
		ds^2 = \left(1 - \frac{2M}{r} + \frac{\varepsilon M^2}{r^2}\right)^{-1} dr^2 + r^2 \alpha^2 d\phi^2 = \left(1 - \frac{2M}{r} + \frac{\varepsilon M^2}{r^2}\right)^{-1} dr^2 + R^2 d\phi^2.\tag{2.8}
	\end{equation}
	Embedding this metric into a flat Euclidean space $(R, z, \phi)$, we have
	\begin{equation}
		d\sigma^2 = dR^2 + dz^2 + R^2 d\phi^2 = \left[ \left(\frac{dR}{dr}\right)^2 + \left(\frac{dz}{dr}\right)^2 \right] dr^2 + R^2 d\phi^2.\tag{2.9}
	\end{equation}
	Comparing the two expressions above yields
	\begin{equation}
		\frac{dz}{dr} = \pm \sqrt{\left(1 - \frac{2M}{r} + \frac{\varepsilon M^2}{r^2}\right)^{-1} - \left(\frac{dR}{dr}\right)^2} = \pm \sqrt{\frac{r^2}{r^2 - 2Mr + \varepsilon M^2} - \alpha^2}.\tag{2.10}
	\end{equation}
    Integrating the above expression gives an analytical solution for $z(r)$; however, given its lengthy form, we omit it here and present the embedding diagram in Fig.~1. It is evident that the embedding surface is compressed along the $z$-axis due to the $-\alpha^2$ under the square root. Similarly, a non-vanishing charge parameter $\varepsilon$ further suppresses the integrand, resulting in additional compression along the same direction.

		\begin{figure}[htbp]
			\centering
			\begin{subfigure}[b]{0.45\textwidth}
				\centering
				\includegraphics[width=\textwidth]{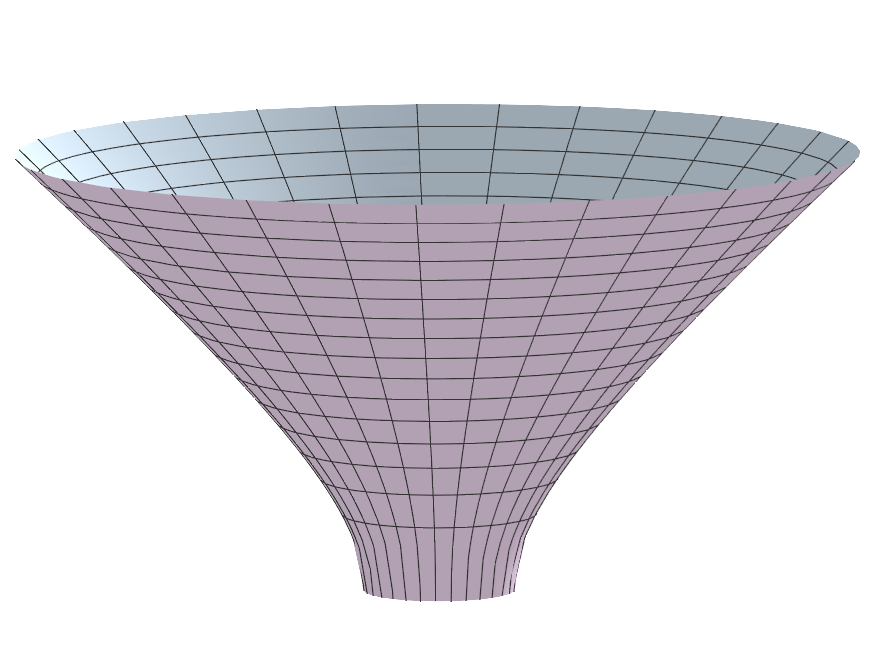}  
				\caption{}
			\end{subfigure}
			\hfill  
			\begin{subfigure}[b]{0.45\textwidth}
				\centering
				\includegraphics[width=\textwidth]{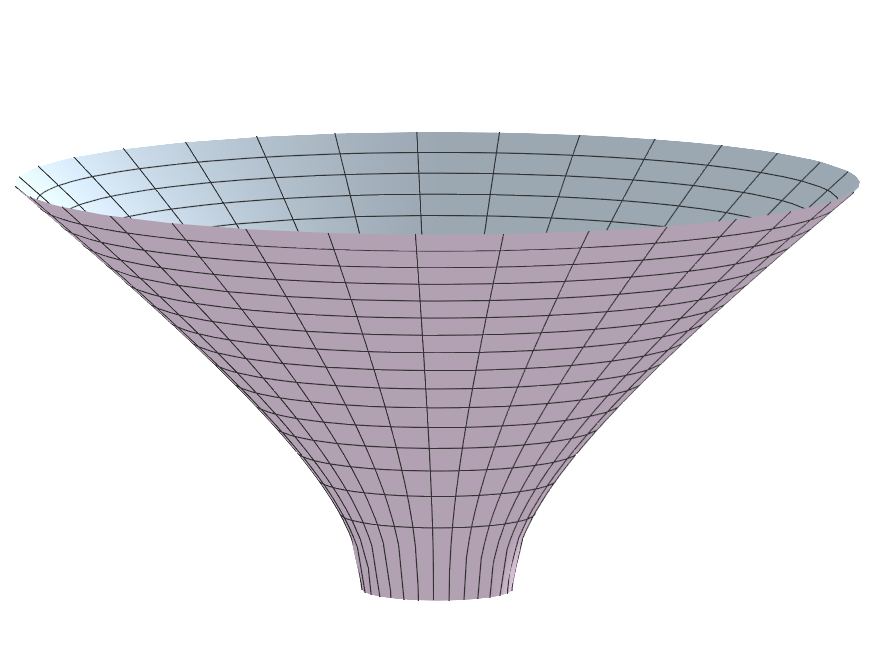}  
				\caption{}
			\end{subfigure}
			\caption*{Fig.1: Embedding diagram for (a) $\alpha = 0$ and (b) $\alpha = 0.5$, with $M = 1$ and $\varepsilon = 0.36$.}  
			\label{fig:two}
		\end{figure}

	\section*{3. Equations of motion and classification of orbits}
	
	To obtain the geodesics in the conical RN spacetime, we first recall its Killing vectors, which yield conserved quantities along the geodesics. According to Ref. [18], the spacetime described by the metric (2.2) admits two Killing vectors, $\xi_{(t)} = \partial/\partial t$ and $\xi_{(\phi)} = \alpha^{-1} (\partial/\partial \phi)$. These respectively give rise to the conservation of energy and of angular momentum about the string axis
	\begin{equation}
		E = -g_{tt}\dot{t} = \left(1 - \frac{2M}{r} + \frac{\varepsilon M^2}{r^2}\right)\dot{t}, \quad L_z = g_{\phi \phi}\alpha^{-1}\dot{\phi} = r^2 \alpha \sin^2\theta \, \dot{\phi}.\tag{3.1}
	\end{equation}
	Here, the dot denotes differentiation with respect to the affine parameter $\eta$. Moreover, the squared magnitude of the angular momentum vector $\vec{L} = (L_x, L_y, L_z)$, denoted $L^2$, is also conserved
	\begin{equation}
		L^2 = \dot{\theta}^2 r^4 + L_z^2 \sin^{-2} \theta.\tag{3.2}
	\end{equation}
	The other two components of the angular momentum are given, respectively, by [18]
    \begin{equation}
	   L_x = -r^2 \sin(\alpha\phi)\dot{\theta} - \alpha r^2 \cos(\alpha\phi)\sin\theta\cos\theta\dot{\phi},\tag{3.3a}
    \end{equation}
	\begin{equation}
		L_y = r^2 \cos(\alpha\phi)\dot{\theta} - \alpha r^2 \sin(\alpha\phi)\sin\theta\cos\theta\dot{\phi}.\tag{3.3b}
	\end{equation}
	
	To derive the equations of motion, we introduce the following Lagrangian: $\mathcal{L} = -\frac{1}{2} g_{\mu\nu}\dot{x}^\mu\dot{x}^\nu$. Since the four-velocity of a photon is null, the Lagrangian vanishes. Combining the Euler-Lagrange equations with the conserved quantities leads to the following equations of motion
	\begin{equation}
		\begin{split}
			\dot{t}^2 &= E^2\left(1 - \frac{2M}{r} + \frac{\varepsilon M^2}{r^2}\right)^{-2}, \\
			\quad \dot{r}^2 &= E^2 - V_{\text{eff}}(r), \\
			\dot{\theta}^2 &= \frac{L^2}{r^4} - \frac{L_z^2}{r^4 \sin^2\theta}, \\
			\quad \dot{\phi}^2 &= \frac{L_z^2}{\alpha^2 r^4 \sin^4\theta},
		\end{split}\tag{3.4}
	\end{equation}
	where the effective potential $V_{\text{eff}}(r)$ is defined as
	\begin{equation}
		V_{\text{eff}}(r) = \left(1 - \frac{2M}{r} + \frac{\varepsilon M^2}{r^2}\right) \frac{L^2}{r^2}.\tag{3.5}
	\end{equation}
	From these equations, two immediate properties follow. First, regarding the radial motion, for a photon outside the event horizon initially moving away from the black hole, when its energy exceeds the local maximum of the effective potential $V_{\text{eff}}^c$, it can overcome the potential barrier and escape to infinity. Second, since $\dot{\theta}^2 \ge 0$, the $\theta$-motion of the photon is confined to the interval
	\begin{equation}
		\theta \in \left[\arcsin\left(\frac{L_z}{L}\right), \pi - \arcsin\left(\frac{L_z}{L}\right)\right],\tag{3.6}
	\end{equation}
	in analogy with the conical Schwarzschild spacetime [20]. In the special case $L_z = L$, both $L_x$ and $L_y$ vanish, confining the motion to the equatorial plane.
	
	Combining $\dot{\theta}^2$ and $\dot{\phi}$ to eliminate the affine parameter yields the differential equation for $\theta$  with respect to $\phi$ as
	\begin{equation}
		\frac{d\theta}{d\phi} = \alpha \sin\theta \sqrt{\left( \frac{L^2}{L_z^2} - 1 \right)\sin^2\theta - 1}.\tag{3.7}
	\end{equation}
	This equation admits the solution
	\begin{equation}
		\cot^2 \theta = \left( \frac{L^2}{L_z^2} - 1 \right) \sin^2(\alpha \phi).\tag{3.8}
	\end{equation}
    Substituting the relation between $\theta$ and $\phi$ obtained above back into the fundamental equations of motion gives the differential relations of $r$ with respect to $\theta$ and $\phi$, respectively
	\begin{equation}
		\left( \frac{dr}{d\theta} \right)^2 = \frac{r^4 \sin^2 \theta}{L^2 \sin^2 \theta - L_z^2} \left[ E^2 - \frac{L^2}{r^2} \left( 1 - \frac{2M}{r} + \frac{\varepsilon M^2}{r^2} \right) \right],\tag{3.9a}
	\end{equation}
	\begin{equation}
		\left( \frac{dr}{d\phi} \right)^2 = \frac{\alpha^2 r^4}{L_z^2 \left[ \left( L^2/L_z^2 - 1 \right) \sin^2(\alpha \phi) + 1 \right]^2} \left[ E^2 - \frac{L^2}{r^2} \left( 1 - \frac{2M}{r} + \frac{\varepsilon M^2}{r^2} \right) \right]. \tag{3.9b}
	\end{equation}
    Our primary focus is on solving the differential equations in (3.9). To simplify the calculation, we introduce the parameters: $\varpi = E^2$, $\lambda = \frac{4M^2}{L^2}$, and a new variable $x = \frac{2M}{r} - \frac{1}{3}$. In terms of these quantities, Eqs.~(3.9) can be recast as
	\begin{equation}
		\frac{dx}{\sqrt{P(x)}} = \frac{1}{2} \left[ 1 - \frac{1}{(L/L_z)^2 \sin^2 \theta} \right]^{-1/2} d\theta, \tag{3.10a}
	\end{equation}
	\begin{equation}
		\frac{dx}{\sqrt{P(x)}} = \frac{1}{2} \frac{\alpha L/L_z}{(L^2/L_z^2 - 1) \sin^2(\alpha \phi) + 1} d\phi, \tag{3.10b}
	\end{equation}
	where the quartic polynomial $P(x)$ reads
	\begin{equation}
		P(x) = -\varepsilon x^4 + 4\left(1 - \frac{\varepsilon}{3}\right)x^3 - \frac{2\varepsilon}{3}x^2 - 4\left(\frac{1}{3} + \frac{\varepsilon}{27}\right)x + 4\left(\varpi \lambda - \frac{\varepsilon}{324} - \frac{2}{27}\right).\tag{3.11}
	\end{equation}
	Obviously, solutions exist only for $P(x) > 0$, allowing us to classify them according to the zeros of the characteristic polynomial $P(x)$. Meanwhile, the requirement $r \geq 0$ imposes the constraint $x \geq -\frac{1}{3}$. Since $P(x)$ is a quartic polynomial, it can have at most four distinct real zeros.  Given the richness of possible root configurations, we analyze them in detail below. A general quartic polynomial
	\begin{equation}
		f(x) = ax^4 + bx^3 + cx^2 + dx + e, \quad (a \neq 0) \nonumber
	\end{equation}
	can be reduced to a depressed form by eliminating the cubic term via the substitution $x = y - \frac{b}{4a}$. Applying this transformation to $P(x)$ yields the depressed quartic
	\begin{equation}
		P(y) = -\varepsilon \left( y^4 + py^2 + qy + s \right), \tag{3.12a}
	\end{equation}
	with coefficients
	\begin{equation}
		p = -\frac{6}{\varepsilon^2} + \frac{4}{\varepsilon}, \quad q = -\frac{8}{\varepsilon^3} + \frac{8}{\varepsilon^2}, \quad s = -\frac{3}{\varepsilon^4} + \frac{4}{\varepsilon^3} - \frac{4\varpi\lambda}{\varepsilon}. \tag{3.12b}
	\end{equation}
	The constraint $x \geq -\frac{1}{3}$ translates to the domain $y \in [-1/\varepsilon, \infty)$. The zeros of $P(y)$ are determined by the quartic equation: $y^4 + py^2 + qy + s = 0$, which is characterized by the discriminant
	\begin{equation}
		\Delta = -4096 \varepsilon^{-6} \chi \left[ 4\varepsilon^3 \chi^2 + \left( 8\varepsilon^2 - 36\varepsilon + 27 \right) \chi + 4(\varepsilon - 1) \right], \tag{3.13}
	\end{equation}
	
	\noindent where $\chi = \varpi \lambda > 0$. The sign of the discriminant $\Delta$ is determined by the quadratic factor in brackets. For $0 < \varepsilon \leq 1$, this quadratic admits two real roots of opposite signs
	\begin{equation}
		\chi_* = \frac{-27 + 36\varepsilon - 8\varepsilon^2 + (9 - 8\varepsilon)^{3/2}}{8\varepsilon^3} > 0, \quad \chi_2 = \frac{-27 + 36\varepsilon - 8\varepsilon^2 - (9 - 8\varepsilon)^{3/2}}{8\varepsilon^3} < 0. \tag{3.14}
	\end{equation}
	Since $\chi > 0$, only $\chi_*$ is physically relevant. Subsequently, a more detailed classification of the zeros of the quartic equation depends on the relationships among $p$, $q$, and $s$. Despite the complexity of the process (detailed in Appendix A), the roots of $P(y)$ can be succinctly classified into three categories according to whether the discriminant $\Delta$ is positive, zero, or negative. Specifically,
	\begin{itemize}
		\item $\Delta = 0$ ($\chi = \chi_*$), the equation has two distinct real roots and one real double root;
		\item $\Delta > 0$ ($0 < \chi < \chi_*$), all four roots are real and distinct;
		\item $\Delta < 0$ ($\chi > \chi_*$), two real roots and a complex conjugate pair.
	\end{itemize}
	Fig.2 displays the profiles of $P(y)$ corresponding to each of these three cases.
	
	\begin{figure}[H]
		\centering
		\includegraphics[width=0.65\textwidth]{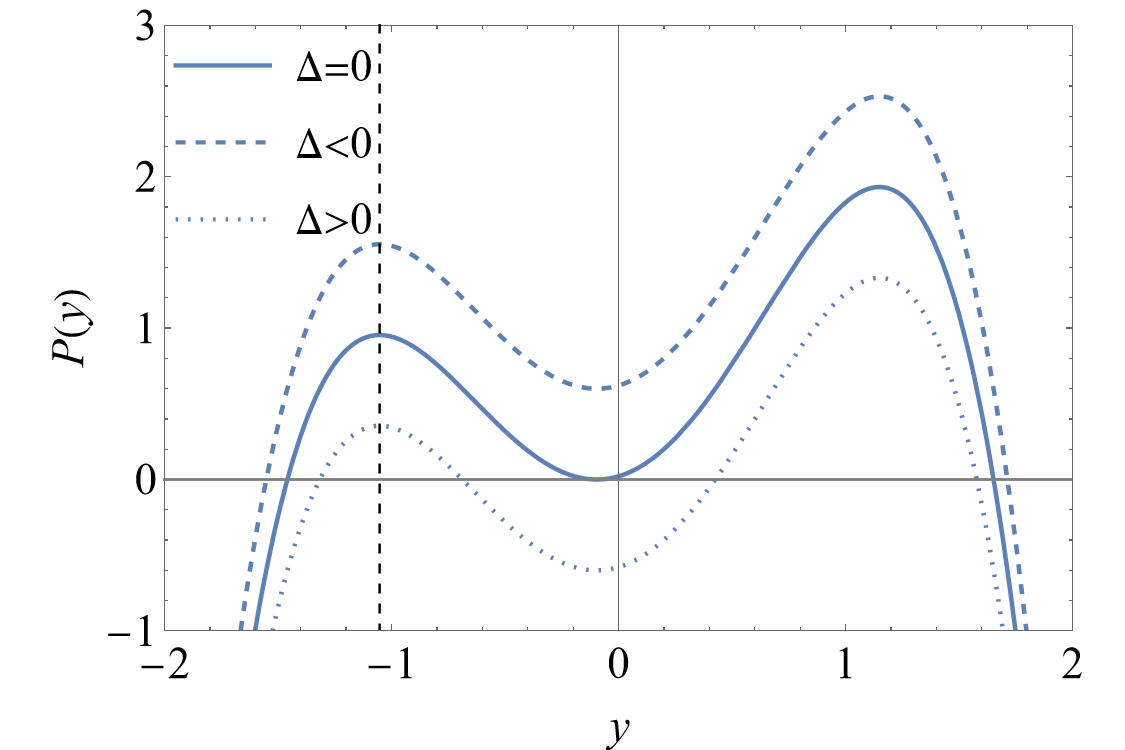}  
		\caption*{Fig.2: The function $P(y)$ for three distinct cases, determined by the sign of the discriminant $\Delta$, where $\varepsilon = 0.95$. The vertical dashed line marks $y = -1/\varepsilon$; the interval to the right of this dashed line is physically meaningful.}
		\label{fig:single}
	\end{figure}
	
	The preceding analysis reveals that the dimensionless parameter $\chi = 4M^2 (E^2/L^2)$, which encodes the ratio of a photon’s energy $E$ to its angular momentum $L$, plays a central role in classifying null geodesics. In fact, the three orbital regimes identified via the critical value $\chi_*$ correspond precisely to distinct ranges of the ratio $E/L$. Since $\chi_*$ is uniquely determined by $\varepsilon$, i.e., $Q^2/M^2$, for a black hole of fixed mass, the charge $Q$ becomes the sole physical parameter governing the threshold $\chi_*$ that demarcates these orbital types. To clearly illustrate the correspondence between the roots of the polynomial $P(x)$ and the null geodesics, we present comparative plots in Fig.3 for the case $\varepsilon = 0.6$. All three cases are plotted with a fixed energy $E$ while varying angular momentum $L$. 
	
	\vspace{1em}
	\noindent Panels (a) and (b) correspond to the case $\chi = \chi_* = 4M^2 (E^2/L^2)$, where $E^2 = V_{\text{eff}}^c(r_0)$. As a reminder, $V_{\text{eff}}^c(r_0)$ represents the critical effective potential (the local maximum of the effective potential), with $r_0$ denoting the radius at which the maximum occurs. In this case, the geodesics fall into three distinct categories: 
	\begin{itemize}
		\item Outside $r_0$, photons arriving from spatial infinity asymptotically approach $r_0$ but cannot surmount the potential barrier.
		\item Inside $r_0$, any ingoing light ray is blocked by a singular potential barrier --- specifically, an infinitely high effective potential barrier. (This singular barrier lies inside the inner horizon, where the radial coordinate $r$ becomes spacelike again. According to the Penrose diagram, light rays could in principle reflect off this barrier and emerge into another universe, though such considerations lie beyond the scope of our present discussion.)
		\item For photons located exactly at $r = r_0$, null geodesics can form constant-radius orbits, but all such orbits are unstable.
	\end{itemize}
	
	\vspace{1em}
	\noindent
	Panels (c) and (d) correspond to the case $\chi > \chi_*$ (equivalently $E^2 > V_{\text{eff}}^c$); photons coming from infinity can overcome the critical potential barrier, cross the event horizon, and proceed toward the singular potential barrier, where they may subsequently be reflected into another universe. Meanwhile, outgoing null geodesics located outside the event horizon can escape to spatial infinity.
	
	\vspace{1em}
	\noindent
	Panels (e) and (f) present the final case, $\chi < \chi_*$ ($E^2 < V_{\text{eff}}^c$), which includes two sub-scenarios:
	\begin{itemize}
		\item For null geodesics originating from spatial infinity, the critical potential barrier reflects them back to infinity before they can reach the event horizon.
	    \item For null geodesics located at $r < r_0$, regardless of their initial direction of motion, they will inevitably encounter the infinitely high potential barrier interior to the inner horizon.
	\end{itemize}
	
	\begin{figure}[H]
		\centering
		
		\begin{subfigure}[b]{0.48\textwidth}
			\centering
			\includegraphics[width=\textwidth]{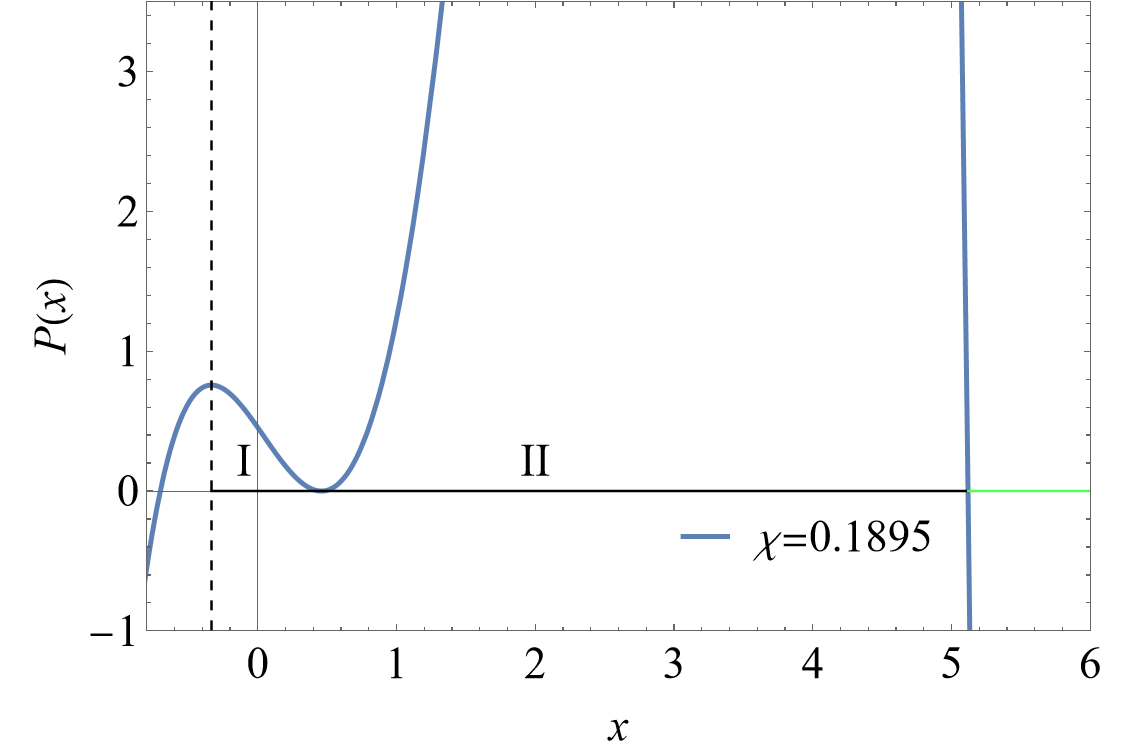} 
			\caption{}
			\label{}
		\end{subfigure}
		\hfill 
		\begin{subfigure}[b]{0.5\textwidth}
			\centering
			\includegraphics[width=\textwidth]{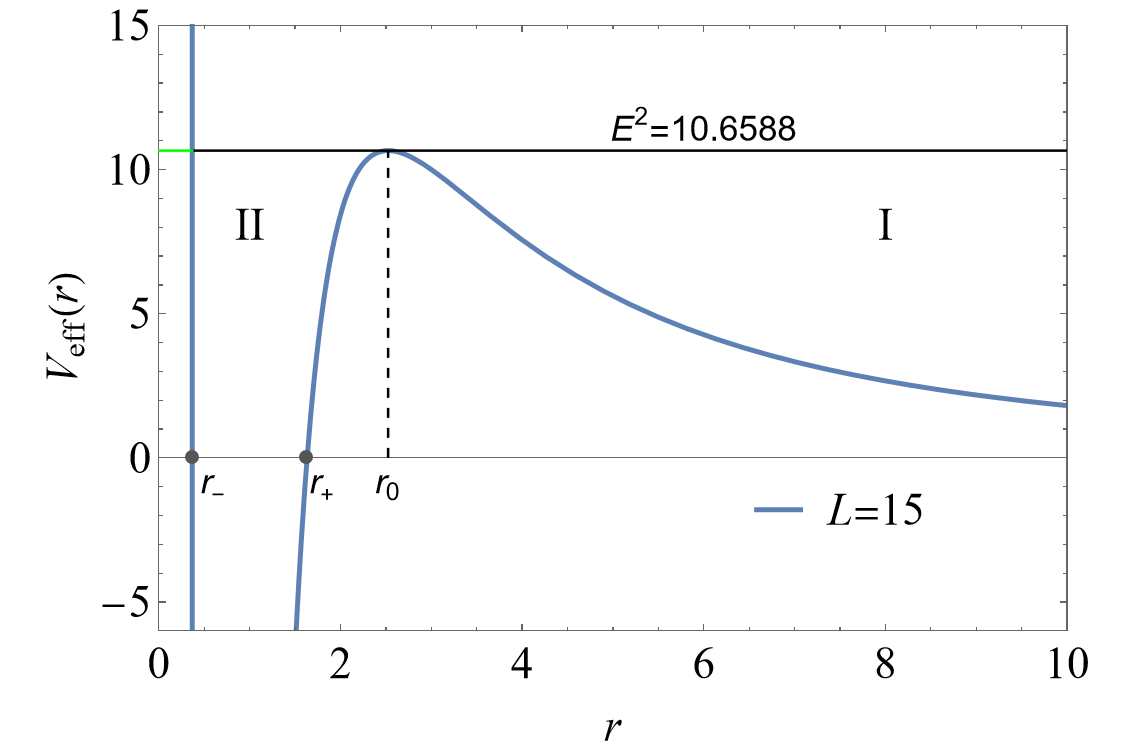}
			\caption{}
			\label{}
		\end{subfigure}
		\end{figure}
		
		\begin{figure}[H] 
			\ContinuedFloat
			\centering
		\begin{subfigure}[b]{0.48\textwidth}
			\centering
			\includegraphics[width=\textwidth]{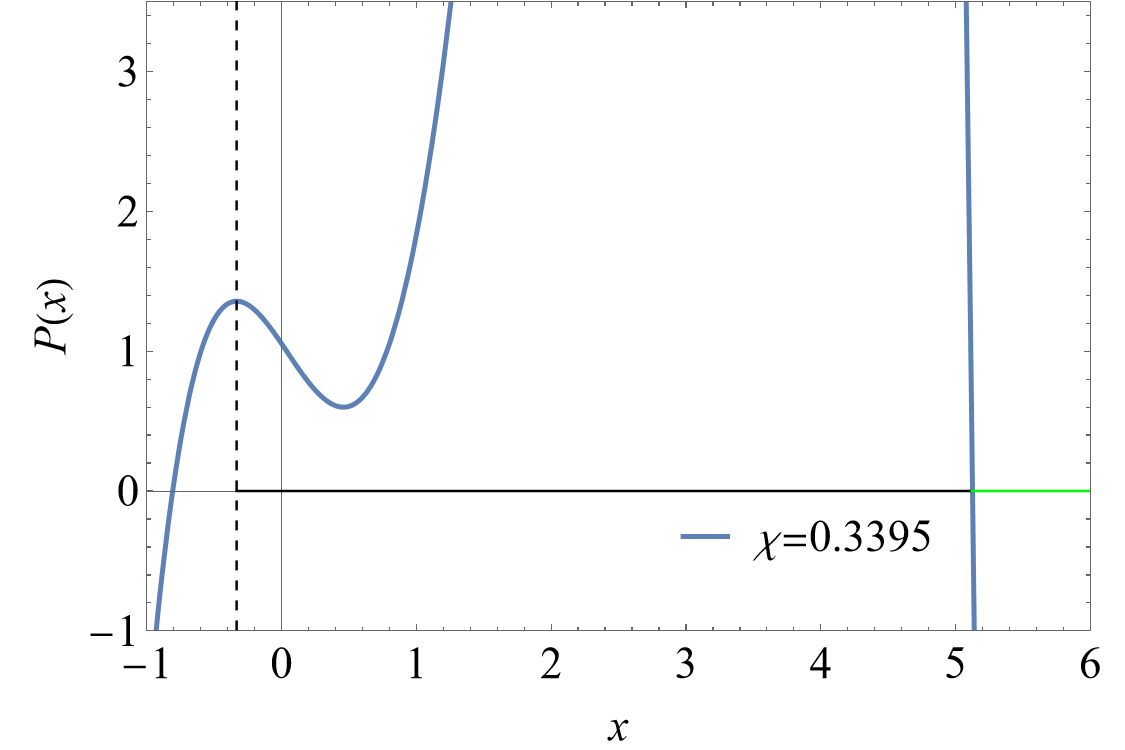}
			\caption{}
			\label{}
		\end{subfigure}
		\hfill
		\begin{subfigure}[b]{0.5\textwidth}
			\centering
			\includegraphics[width=\textwidth]{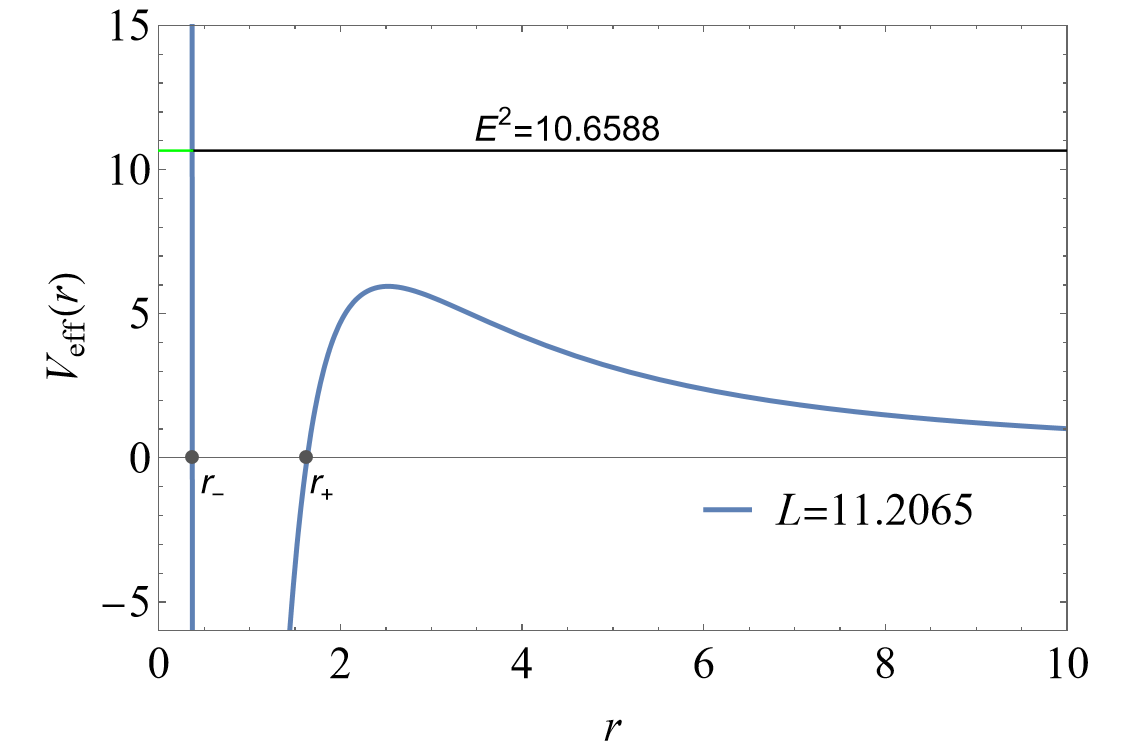}
			\caption{}
			\label{}
		\end{subfigure}
	\end{figure}
	
	\begin{figure}[H] 
		\ContinuedFloat
		\centering
		\vspace{-1.5em}
		\begin{subfigure}[b]{0.48\textwidth}
			\centering
			\includegraphics[width=\textwidth]{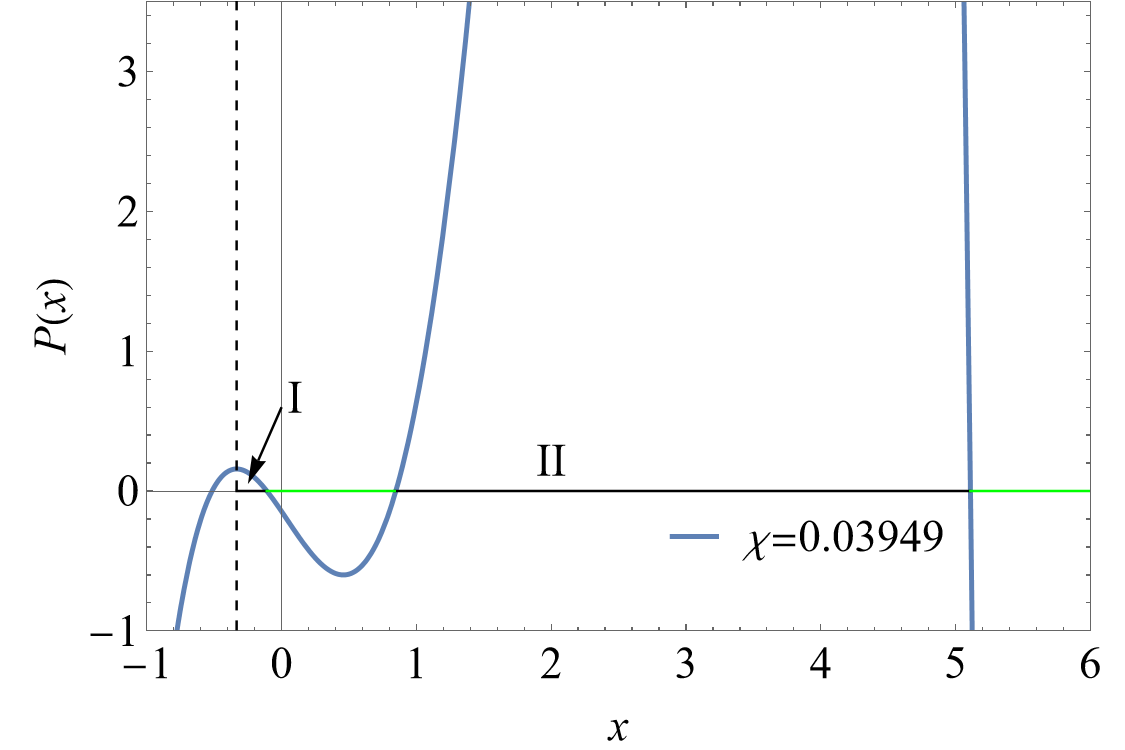}
			\caption{}
			\label{}
		\end{subfigure}
		\hfill
		\begin{subfigure}[b]{0.5\textwidth}
			\centering
			\includegraphics[width=\textwidth]{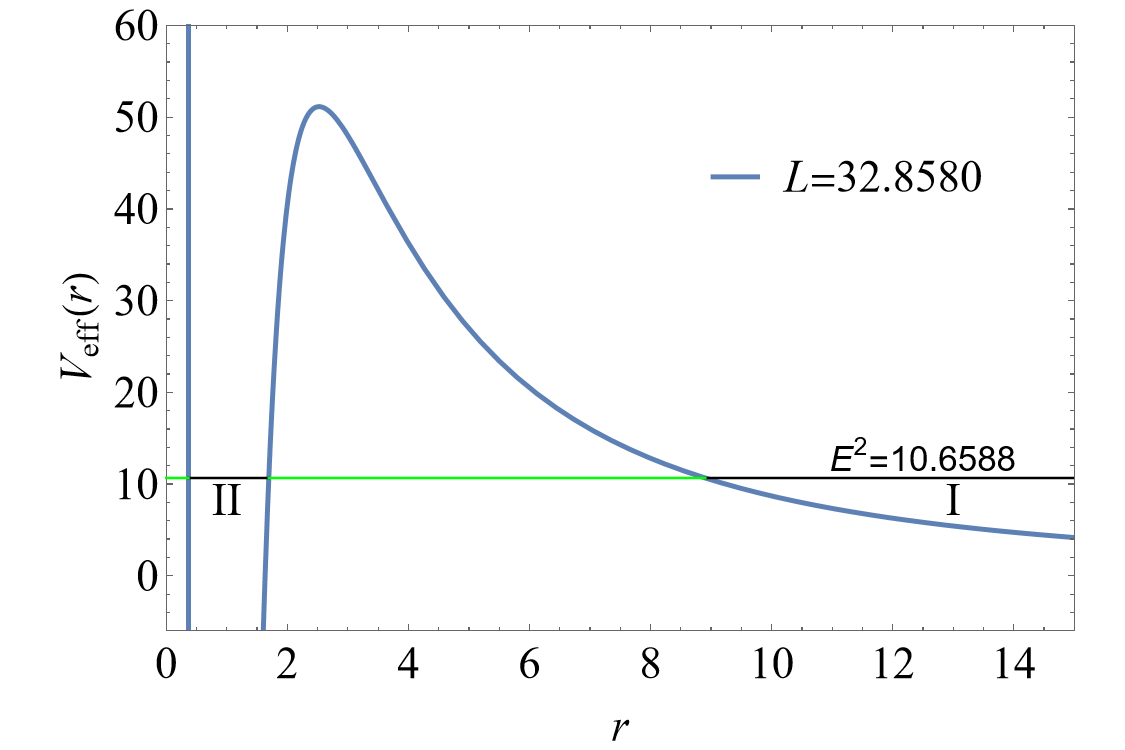}
			\caption{}
			\label{}
		\end{subfigure}
		
		\caption*{Fig.3: For $\varepsilon = 0.6$ (corresponding to $\chi_* = 0.1895$), panels (a), (c), (e) plot $P(x)$ for $\chi = \chi_*$, $\chi > \chi_*$, and $\chi < \chi_*$, respectively, with the corresponding $V_{\text{eff}}(r)$ shown in each row. The green interval indicates the inaccessible region. Here $E^2 = 10.6588$ is fixed throughout.}
		\label{fig:total}
	\end{figure}
	
	\section*{4. Solution of the geodesic equations}
	
	The previous section classified null geodesics into three main categories by analyzing the roots of the polynomial $P(x)$ and provided a qualitative description for each type. This section derives analytical solutions for each category by integrating Eqs. (3.10a) and (3.10b), which respectively yield
	\begin{equation}
		\int_{x(\theta_0)}^{x(\theta)} \frac{dx}{\sqrt{P(x)}} = \mathcal{F}(\theta) = \frac{1}{2} \left[ \arcsin \left( \frac{\cos \theta_0}{\sqrt{1 - L_z^2/L^2}} \right) - \arcsin \left( \frac{\cos \theta}{\sqrt{1 - L_z^2/L^2}} \right) \right], \tag{4.1a}
	\end{equation}
	\begin{equation}
		\int_{x(\phi_0)}^{x(\phi)} \frac{dx}{\sqrt{P(x)}} = \mathcal{G}(\phi) = \frac{1}{2} \left\{ \arctan \left[ \frac{L}{L_z} \tan(\alpha \phi) \right] - \arctan \left[ \frac{L}{L_z} \tan(\alpha \phi_0) \right] \right\}. \tag{4.1b}
	\end{equation}
	Rotational symmetry allows us to set the initial condition $\phi_0 = 0$ without loss of generality, leading to $\mathcal{G}(\phi) = \frac{1}{2} \arctan \left[ \frac{L}{L_z} \tan(\alpha \phi) \right]$. This function is piecewise discontinuous, violating the continuity required for null geodesics. To resolve the discontinuity, we regularize it into a continuous, monotonically increasing function
	\begin{equation*}
		\hat{\mathcal{G}}(\phi) = \frac{1}{2} \left\{ \arctan \left[ \frac{L}{L_z} \tan(\alpha \phi) \right] + \pi \left\lfloor \frac{\alpha \phi}{\pi} + \frac{1}{2} \right\rfloor \right\},
	\end{equation*}
	where $\lfloor \cdot \rfloor$ denotes the floor function. However, this adjusted function fails to be differentiable at the jump points due to the presence of the floor function. To remove this non‑differentiability, we apply the identity
	\begin{equation*}
	\arctan\!\left( a \tan x \right) + \pi \left\lfloor \frac{x}{\pi} + \frac{1}{2} \right\rfloor
	= x + \arctan\!\left[ \frac{(a - 1) \sin 2x}{(a + 1) - (a - 1) \cos 2x} \right],
    \end{equation*}
	which yields a smooth, differentiable expression
	\begin{equation}
	\tilde{\mathcal{G}}(\phi) = \frac{1}{2} \left\{ \alpha\phi + \arctan\!\left[ \frac{(L/L_z - 1) \sin(2\alpha\phi)}{(L/L_z + 1) - (L/L_z - 1) \cos(2\alpha\phi)} \right] \right\}.\tag{4.2}
	\end{equation}
	The validity of the above transformations is confirmed by noting that the derivative with respect to $\phi$ yields the same original function. Regarding the $\theta$-motion, we choose $\theta_0 = \pi - \arcsin(L_z/L)$, i.e., the null geodesic starts from the turning point of the $\theta$ coordinate. With this choice, $\mathcal{F}(\theta) = -\frac{1}{2} \left[ \arcsin \left( \frac{\cos \theta}{\sqrt{1 - L_z^2/L^2}} \right) + \frac{\pi}{2} \right]$. The integral on the left‑hand side of Eq.~(4.1) will be treated separately for two distinct cases below.
	
    \subsection*{4.1 The case $\Delta = 0$}
	
	\hspace{1.5em}For the case $\Delta = 0$, $P(x)$ possesses one real double root and two distinct real roots, allowing it to be factored as
	\begin{equation*}
		P(x) = -\varepsilon (x - x_1) (x - x_2)^2 (x - x_3),
	\end{equation*}
	where $x_1 < -1/3 < x_2 < x_3$ are the zeros of $P(x)$. According to Fig.1 (recall that $P(x)$ and $P(y)$ differ only by a horizontal shift), the integration domain naturally splits into two parts. For the interval between $-1/3$ and $x_2$, substituting the factored form of $P(x)$ yields
    \begin{equation}
	\begin{split}
		&\int_{x_\text{init}}^{x_\text{f}} \frac{dx}{\sqrt{P(x)}}= \int_{x_\text{init}}^{x_\text{f}} \frac{1}{\sqrt{\varepsilon}(x_2-x)\sqrt{(x-x_1)(x_3-x)}} \\
		&= \frac{1}{\sqrt{\varepsilon}\sqrt{(x_2-x_1)(x_3-x_2)}} \left( \ln \left[ \frac{\sqrt{(x_2-x_1)(x_3-x)} + \sqrt{(x-x_1)(x_3-x_2)}}{\sqrt{(x_2-x_1)(x_3-x)} - \sqrt{(x-x_1)(x_3-x_2)}} \right] \right)  \Bigg|_{x_\text{init}}^{x_\text{f}},
	\end{split}\tag{4.3a}
    \end{equation}
	where $x_{\text{init}}$ and $x_{\text{f}}$ represent the initial and final positions of the null geodesic. When the interval of integration lies between $x_2$ and $x_3$, a similar integration procedure yields
    \begin{equation} 
    	\begin{split}
    		&\int_{x_\text{init}}^{x_\text{f}} \frac{dx}{\sqrt{P(x)}} = \int_{x_\text{init}}^{x_\text{f}} \frac{dx}{\sqrt{\varepsilon}(x-x_2)\sqrt{(x-x_1)(x_3-x)}} \\
    		&= \frac{1}{\sqrt{\varepsilon}\sqrt{(x_2-x_1)(x_3-x_2)}} \left( \ln \left[ \frac{\sqrt{(x_2-x_1)(x_3-x)} - \sqrt{(x-x_1)(x_3-x_2)}}{\sqrt{(x_2-x_1)(x_3-x)} + \sqrt{(x-x_1)(x_3-x_2)}} \right] \right) \Bigg|_{x_\text{init}}^{x_\text{f}}.
    	\end{split}\tag{4.3b}
    \end{equation}
	The implicit relations $r(\theta)$ and $r(\phi)$ follow from substituting the integral result from Eq.~(4.3) into Eq.~(4.1).
	
	The third case corresponds to null geodesics that form constant-radius orbits at $x = x_2$ (corresponding to $r = r_0$). From Eq.~(3.6), solving for $\sin^2\theta$ gives
	\begin{equation}
		\sin^2\theta = \frac{1}{1 + \left( L^2/L_z^2 - 1 \right) \sin^2(\alpha\phi)}.
		\tag{4.4}
	\end{equation}
	Since $\sin\theta$ is non‑monotonic on $[0, \pi]$, we use $\cos^2\theta$ to obtain a one‑to‑one mapping between $\phi$ and $\theta$
	\begin{equation}
		\cos^2\theta = \frac{\left( L^2/L_z^2 - 1 \right) \sin^2(\alpha\phi)}{1 + \left( L^2/L_z^2 - 1 \right) \sin^2(\alpha\phi)}.
		\tag{4.5}
	\end{equation}
	Taking the square root leads to the explicit expression linking $\theta$ and $\phi$
	\begin{equation}
		\theta_\pm(\phi) = \arccos\left[ \pm \frac{\sqrt{ L^2/L_z^2 - 1 } \sin(\alpha\phi)}{\sqrt{1 + \left( L^2/L_z^2 - 1 \right) \sin^2(\alpha\phi)}} \right].
		\tag{4.6}
	\end{equation}
	The two branches are symmetric about the equatorial plane, satisfying $\theta_+ = \pi - \theta_-$.
	
	\subsection*{4.2 The case $\Delta \neq 0$}
	
	\hspace{1.5em}When $\Delta \neq 0$, the quartic polynomial $P(x)$ possesses no repeated roots, rendering the integral $\int \frac{dx}{\sqrt{P(x)}}$ non-elementary. To evaluate it, we employ the transformation
	\begin{equation}
		t (x) = \frac{1}{24} P''(x_0) + \frac{P'(x_0)}{4(x - x_0)},
		\tag{4.7}
	\end{equation}
	where $x_0$ is a real root of $P(x)$. This converts the original integral into the standard Weierstrass form
	\begin{equation}
		z(x) = \int_{x_0}^x \frac{1}{\sqrt{P(x')}} dx' = \int_{t(x)}^\infty \frac{1}{\sqrt{4t'^3 - g_2 t' - g_3}} dt',
		\tag{4.8}
	\end{equation}
	with invariants
	\begin{equation}
		g_2 = 4 \left( \frac{1}{3} - \varepsilon \varpi \lambda \right), \quad
		g_3 = 4 \left[ \frac{2}{27} + \frac{1}{3} (-3 + 2\varepsilon) \varpi \lambda \right].
		\tag{4.9}
	\end{equation}
	The integral result $z(x)$ can be compactly expressed via the inverse Weierstrass elliptic function. Using the definition $\wp^{-1}(t) = \int_{\infty}^t \frac{dt'}{\sqrt{4t'^3 - g_2 t' - g_3}}$, we obtain
	\begin{equation}
		z(x) = -\wp^{-1} \left( \frac{1}{24} P''(x_0) + \frac{P'(x_0)}{4(x - x_0)}; g_2, g_3 \right).
		\tag{4.10}
	\end{equation}
	In the following, we use this result to derive analytical solutions separately for $\Delta < 0$ and $\Delta > 0$.
	
    \subsubsection*{(a) $\Delta < 0$}
	 \hspace{1.5em}When $\Delta < 0$, the polynomial $P(x)$ has two real roots lying on opposite sides of $-1/3$, denoted $x_1 < -1/3 < x_2$. For null geodesics approaching the black hole from $x_\text{init}$, the analytical solution reads
     \begin{equation}
		\int_{x_\text{init}}^{x_\text{f}} \frac{dx}{\sqrt{P(x)}} = \int_{x_1}^{x_\text{f}} \frac{dx}{\sqrt{P(x)}} - \int_{x_1}^{x_\text{init}} \frac{dx}{\sqrt{P(x)}} = \wp^{-1} \left( \frac{1}{24} P''(x_1) + \frac{P'(x_1)}{4(x - x_1)}; g_2, g_3 \right) \bigg|_{x_\text{f}}^{x_\text{init}}.
		\tag{4.11}
	\end{equation}
	While Eq. (4.11) provides a formal solution, it is often more convenient in physics to express the integral in terms of the standard elliptic integral. In this regime, the cubic equation $4t'^3 - g_2 t' - g_3 = 0$ possesses one real root $e_1$ and a pair of complex conjugate roots $e_2 = \alpha + i\beta$ and $e_3 = \alpha - i\beta$ (with $\beta > 0$). Introducing the auxiliary quantities
	\begin{equation}
		A = \sqrt{(e_1 - \alpha)^2 + \beta^2}, \quad k^2 = \frac{A + \alpha - e_1}{2A}, \quad g = \frac{1}{2\sqrt{A}},
		\tag{4.12}
	\end{equation}
	and applying (241.00) from Ref. [30], the integral in Eq. (4.8) transforms into
	\begin{equation}
		\int_{y}^{\infty} \frac{dt}{\sqrt{4t^3 - g_2 t - g_3}} = g F(\varphi, k), \quad \left( \varphi = \arccos\left( \frac{y - e_1 - A}{y - e_1 + A} \right), \ y \ge e_1 \right).
		\tag{4.13}
	\end{equation}
	Here, $F(\varphi, k) = \int_0^\varphi (1 - k^2 \sin^2 \theta)^{-1/2} \, d\theta$ denotes the elliptic integral of the first kind. Consequently, the integral in Eq. (4.8) becomes
	\begin{equation}
		\int_{x_0}^{x} \frac{1}{\sqrt{P(x')}} dx' = \int_{t(x)}^{\infty} \frac{1}{\sqrt{4t'^3 - g_2 t' - g_3}} dt' = g F(\varphi, k), \quad \varphi = \arccos \left[ \frac{t(x) - e_1 - A}{t(x) - e_1 + A} \right].
		\tag{4.14}
	\end{equation}
	Finally, the definite integral from $x_\text{init}$ to $x_\text{f}$ is given by
	\begin{equation}
		\begin{split}
			\int_{x_\text{init}}^{x_\text{f}} \frac{dx}{\sqrt{P(x)}} &= \int_{x_1}^{x_\text{f}} \frac{dx}{\sqrt{P(x)}} - \int_{x_1}^{x_\text{init}} \frac{dx}{\sqrt{P(x)}} \\
			&= \int_{t({x_\text{f}})}^{\infty} \frac{dt}{\sqrt{4t^3 - g_2 t - g_3}} - \int_{t({x_\text{init}})}^{\infty} \frac{dt}{\sqrt{4t^3 - g_2 t - g_3}} \\
			&= g \big[ F(\varphi_2, k) - F(\varphi_1, k) \big],
		\end{split}
		\tag{4.15}
	\end{equation}
	where
	\begin{equation}
		\varphi_1 = \arccos \left[ \frac{t({x_\text{init}}) - e_1 - A}{t({x_\text{init}}) - e_1 + A} \right], \quad \varphi_2 = \arccos \left[ \frac{t({x_\text{f}}) - e_1 - A}{t({x_\text{f}}) - e_1 + A} \right].
		\tag{4.16}
	\end{equation}
	Thus, we have obtained an equivalent expression for Eq.~(4.11) in terms of elliptic integrals.
	
	\subsubsection*{(b) $\Delta > 0$} 
	\hspace{1.5em}When $\Delta > 0$, $P(x) = 0$ admits four distinct real roots. As illustrated in Fig.3, we order them as $x_1 < -1/3 < x_2 < x_3 < x_4$. As previously discussed, the domain of integration splits into two physically relevant intervals. For the interval between $-1/3$ and $x_2$, the integral can be expressed via the inverse Weierstrass elliptic function, yielding a form identical to Eq.~(4.11). Setting $x_\text{init} = -1/3$ corresponds to null geodesics incoming from infinity and approaching the black hole. For the interval between $x_3$ and $x_4$, the integral takes the form
	\begin{equation}
		\int_{x_\text{init}}^{x_\text{f}} \frac{dx}{\sqrt{P(x)}} = \int_{x_3}^{x_\text{f}} \frac{dx}{\sqrt{P(x)}} - \int_{x_3}^{x_\text{init}} \frac{dx}{\sqrt{P(x)}} = \wp^{-1} \left( \frac{1}{24} P''(x_3) + \frac{P'(x_3)}{4(x - x_3)}; g_2, g_3 \right) \bigg|_{x_\text{f}}^{x_\text{init}}.
		\tag{4.17}
	\end{equation}
	Case (b) is particularly crucial for investigating the observational effects of spacetime; therefore, we also convert these integrals into the standard elliptic integral of the first kind. For the interval from $-1/3$ to $x_2$, although the Weierstrass form remains the same as in case (a), the cubic equation $4t^3 - g_2 t - g_3 = 0$ now possesses three distinct real roots: $e_1 > e_2 > e_3$. In this scenario, the auxiliary quantities are redefined as
	\begin{equation}
		k = \sqrt{\frac{e_2 - e_3}{e_1 - e_3}}, \quad g = \frac{1}{\sqrt{e_1 - e_3}}.
		\tag{4.18}
	\end{equation}
	Applying equation (238.00) in [30] gives
	\begin{equation}
		\int_{y}^{\infty} \frac{dt}{\sqrt{4t^3 - g_2 t - g_3}} = g F(\varphi, k), \quad \left( \varphi = \arcsin \sqrt{\frac{e_1 - e_3}{y - e_3}}, \quad y \ge e_1 \right).
		\tag{4.19}
	\end{equation}
	Noting that $\lim_{x \to x_1} t(x) \to \infty$ and $t(x) \ge t(x_2) = e_1$, the definite integral becomes
	\begin{equation}
		\begin{split}
			\int_{x_\text{init}}^{x_\text{f}} \frac{dx}{\sqrt{P(x)}} &= \int_{x_1}^{x_\text{f}} \frac{dx}{\sqrt{P(x)}} - \int_{x_1}^{x_\text{init}} \frac{dx}{\sqrt{P(x)}} \\
			&= \int_{t(x_\text{f})}^{\infty} \frac{dt}{\sqrt{4t^3 - g_2 t - g_3}} - \int_{t(x_\text{init})}^{\infty} \frac{dt}{\sqrt{4t^3 - g_2 t - g_3}} \\
			&= g \left[ F(\varphi_2, k) - F(\varphi_1, k) \right],
		\end{split}
		\tag{4.20}
	\end{equation}
	where
	\begin{equation}
		\varphi_1 = \arcsin \sqrt{\frac{e_1 - e_3}{t(x_\text{init}) - e_3}}, \quad \varphi_2 = \arcsin \sqrt{\frac{e_1 - e_3}{t(x_\text{f}) - e_3}}.
		\tag{4.21}
	\end{equation}
	For the interval between $x_3$ and $x_4$, employing a similar method and equation (233.00) from [30] yields
	\begin{equation}
			\int_{x_\text{init}}^{x_\text{f}} \frac{dx}{\sqrt{P(x)}}
			= g \big[ F(\varphi_1, k) - F(\varphi_2, k) \big],
		\tag{4.22}
	\end{equation}
	with
	\begin{equation}
		\varphi_1 = \arcsin \sqrt{\frac{t(x_\text{init}) - e_3}{e_2 - e_3}}, \quad
		\varphi_2 = \arcsin \sqrt{\frac{t(x_\text{f}) - e_3}{e_2 - e_3}}.
		\tag{4.23}
	\end{equation}
	Thus, analytical solutions have been obtained for all physically relevant cases, expressible either through the inverse Weierstrass elliptic function or in the standard form of the elliptic integral of the first kind. For clarity, both equivalent representations are summarized in Appendix B.
	
    \section*{5. Numerical Simulation of the Solutions}
    
    In this section, we utilize the analytical solutions to numerically simulate and explore the properties of null geodesics. Similar to the previous sections, the discussion is divided into different cases.
    
    \subsection*{5.1 The case $\Delta = 0$}
    
    \subsubsection*{(a) Orbits approaching the black hole from afar}
    
    \hspace{1.5em}Fig.4 (a) presents a three-dimensional plot of null geodesics approaching the black hole from spatial infinity, with $\alpha = 0.6$. As the null geodesic approaches the black hole, its trajectory begins to wind around it, forming a spiral orbit. This behavior is further corroborated by panel (e), which shows that the polar angle $\theta$ remains confined within the interval $[0.878, 2.264]$, in full agreement with the analytical result derived from Eq.~(3.6). Moreover, our solution (4.1) reveals a novel feature: the angular coordinates of the null geodesics exhibit a recurrence—they return to their initial angular position after completing an azimuthal rotation of $\delta\phi$ around the cosmic string. This phenomenon is clearly visible in Fig.4 (e), and the precise value of $\delta\phi$ will be discussed in detail later. Panel (f) shows that the radial coordinate of the null geodesic asymptotically approaches $r_0$, the radius corresponding to the critical effective potential. Since $x_2$ (corresponding to $r_0$) is a divergent point of the integral in (4.3a), the null geodesic cannot reach $x_2$ within a finite interval of the affine parameter $\eta$. This conclusion can be demonstrated more directly from the equation of motion (2.3). Expanding $V_{\text{eff}}(r)$ around $r_0$ yields
    \begin{equation}
    	V_{\text{eff}}(r) = V_{\text{eff}}(r_0) + \frac{1}{2} V_{\text{eff}}''(r_0) (r-r_0)^2 + \mathcal{O}\bigl((r-r_0)^3\bigr), \tag{5.1}
    \end{equation}
    where $V'_{\text{eff}}(r_0) = 0$ has been used. Noting that $V''_{\text{eff}}(r_0) < 0$ and neglecting higher-order terms, the equation of motion can be written as $\frac{dr}{d\eta} = -\frac{1}{2} \sqrt{-2V''_{\text{eff}}(r_0)} (r-r_0),$
    the solution of which is $ r = r_0 + e^{-\frac{1}{2}\sqrt{-2V''_{\text{eff}}(r_0)} \eta + C_0}.$ As $\eta \to \infty$, $r \to r_0$, indicating that the photon requires an infinite affine parameter to reach $r_0$. Finally, a comparison between panels (a) and (b) demonstrates that for smaller values of $\alpha$ (corresponding to a larger deficit angle), the null geodesic executes more orbital windings as it spirals toward $r_0$. This highlights the crucial role of the deficit angle in shaping the dynamical behavior of null geodesics.
    
     \begin{figure}[H]
    	\centering
    	
    	\begin{subfigure}[b]{0.48\textwidth}
    		\centering
    		\includegraphics[width=\textwidth]{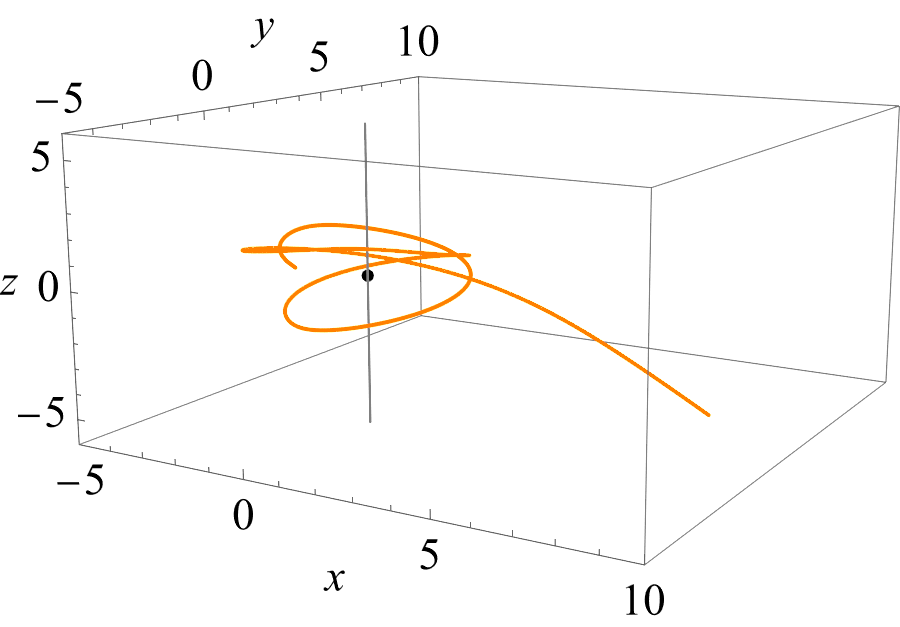} 
    		\caption{}
    		\label{}
    	\end{subfigure}
    	\hfill 
    	\begin{subfigure}[b]{0.48\textwidth}
    		\centering
    		\includegraphics[width=\textwidth]{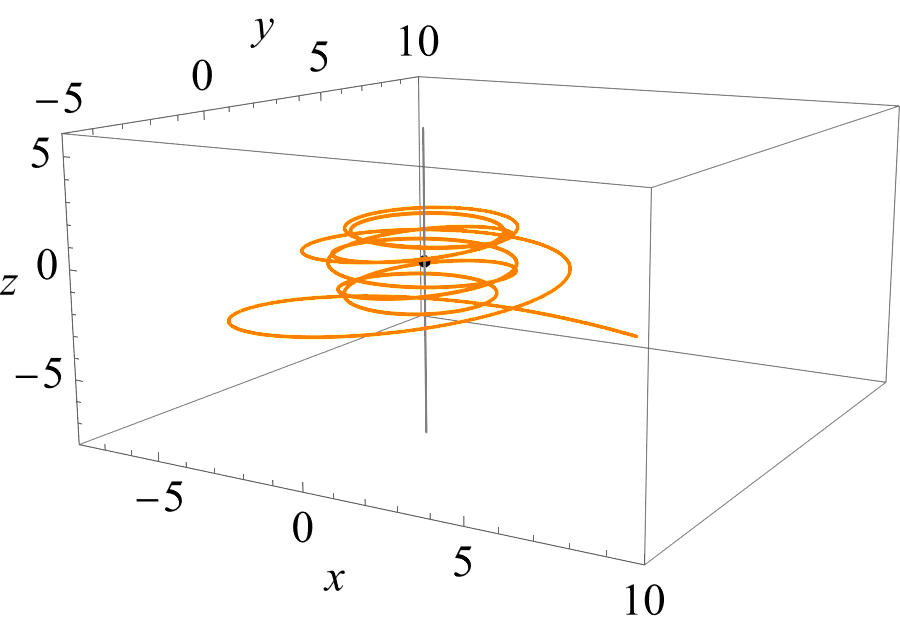}
    		\caption{}
    		\label{}
    	\end{subfigure}
    \end{figure}
    
    \begin{figure}[H] 
    	\ContinuedFloat
    	\centering
    	\begin{subfigure}[b]{0.48\textwidth}
    		\centering
    		\includegraphics[width=\textwidth]{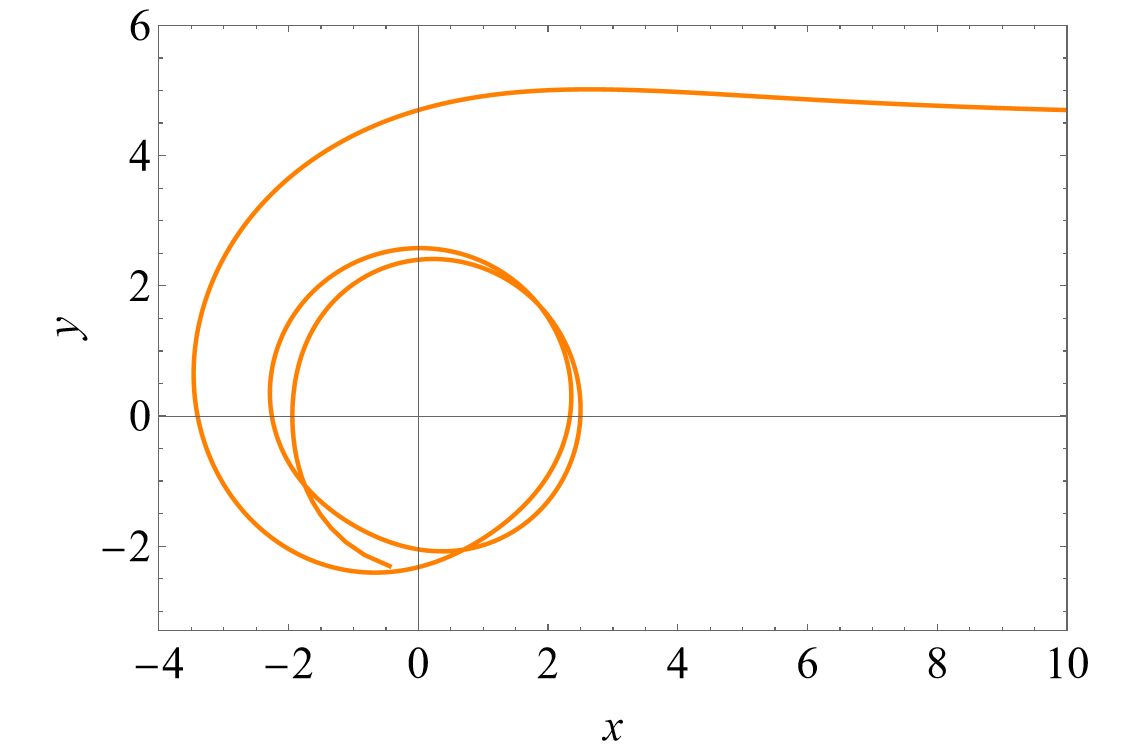}
    		\caption{}
    		\label{}
    	\end{subfigure}
    	\hfill
    	\begin{subfigure}[b]{0.48\textwidth}
    		\centering
    		\includegraphics[width=\textwidth]{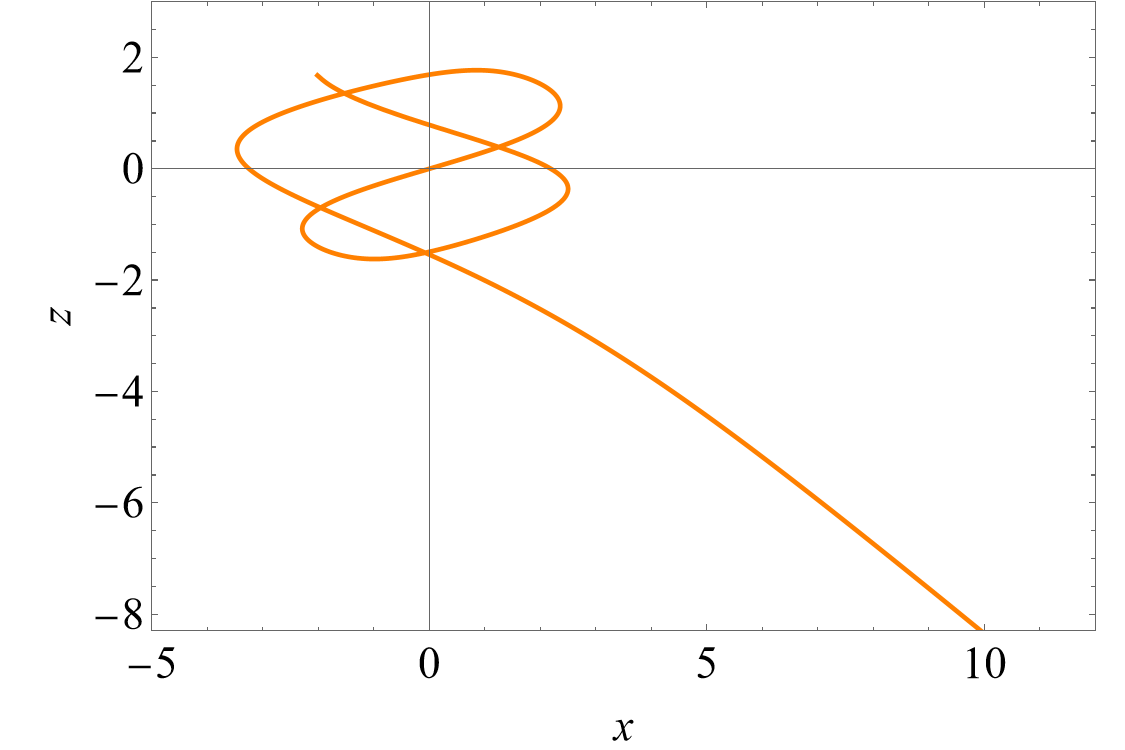}
    		\caption{}
    		\label{}
    	\end{subfigure}
    \end{figure}
    
    \begin{figure}[H] 
    	\ContinuedFloat
    	\centering
    	\vspace{-1.5em}
    	\begin{subfigure}[b]{0.48\textwidth}
    		\centering
    		\includegraphics[width=\textwidth]{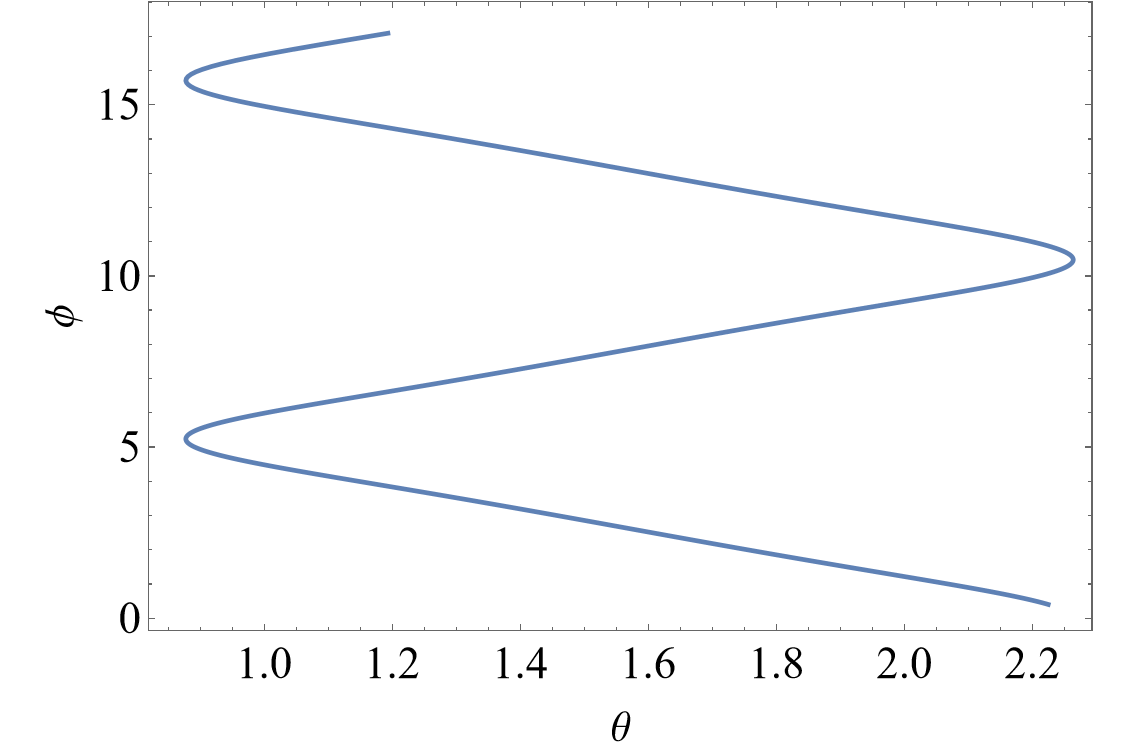}
    		\caption{}
    		\label{}
    	\end{subfigure}
    	\hfill
    	\begin{subfigure}[b]{0.48\textwidth}
    		\centering
    		\includegraphics[width=\textwidth]{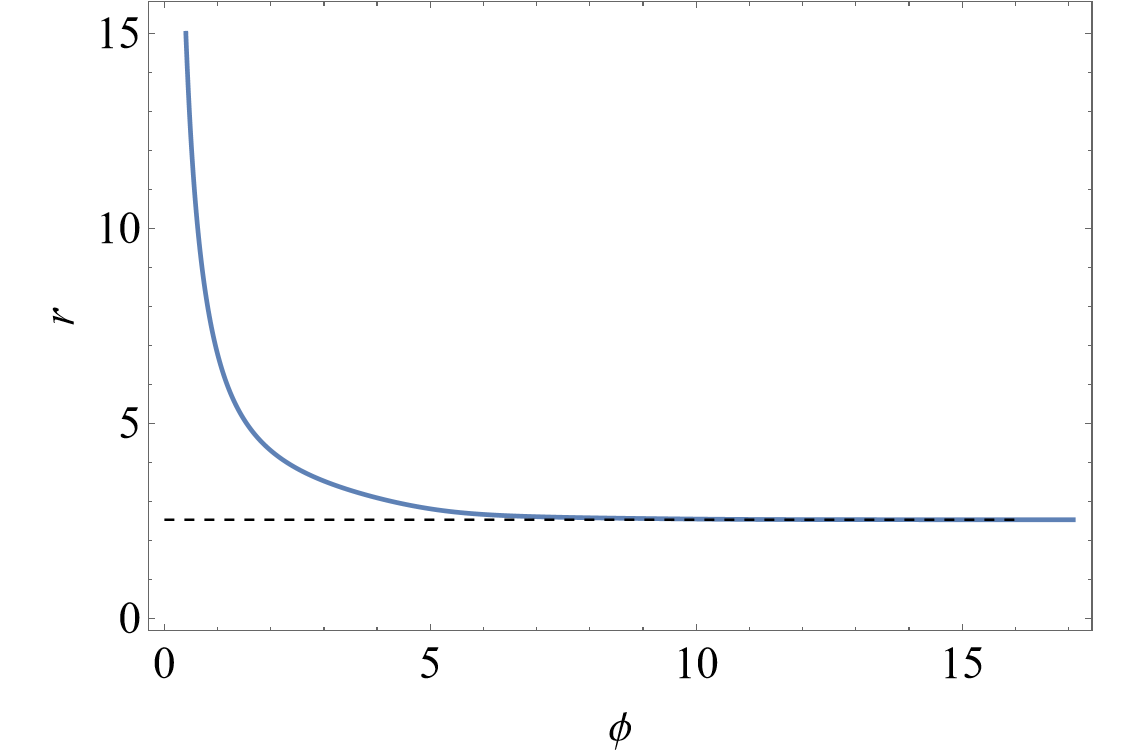}
    		\caption{}
    		\label{}
    	\end{subfigure}
    	
    	\caption*{Fig.4: Simulation results of light rays from infinity approaching a black hole for $\Delta = 0$ ($\chi = \chi_1$), with (a) $\alpha = 0.6$ and (b) $\alpha = 0.2$; panels (c) and (d) present the top view and side view of (a), corresponding to the $x$-$y$ plane (along the cosmic string) and $x$-$z$ plane, respectively; panel (e) illustrates the $\theta-\phi$ evolution for case (a), with $\theta \in [0.878, 2.264]$; panel (f) depicts the corresponding evolution of $r$ as a function of $\phi$, with the dashed line marking $r = r_0$. The parameters are set to $L/L_z = 1.3$ and $\varepsilon = 0.6$ throughout the figure.}
    	\label{fig:total}
    \end{figure}

    We now turn to the influence of the dimensionless charge parameter $\varepsilon$ on null geodesics. Solving $V'_{\text{eff}}(r) = 0$ yields the critical radius $r_0 = \frac{1}{2} M \left( 3 + \sqrt{9 - 8\varepsilon}\right),$ which evidently decreases as $\varepsilon$ increases, allowing null geodesics to approach the origin more closely. The corresponding critical effective potential is given by
    \begin{equation}
    	V_{\text{eff}}^c = \frac{8 L^2 \left( 3 + \sqrt{9 - 8\varepsilon} - 2\varepsilon \right)}{M^2 \left( 3 + \sqrt{9 - 8\varepsilon} \right)^4}.\tag{5.2}
    \end{equation}
    Direct calculation reveals that $V_{\text{eff}}^c$ is a monotonically increasing function of $\varepsilon$ over the physically admissible interval. Consequently, under the condition $\Delta = 0$ (where $V_{\text{eff}}^c = E^2$), the energy $E$ must also increase monotonically with $\varepsilon$. This conclusion is further corroborated by the expression for $\chi_*$. Since $\chi_*$ is an increasing function of $\varepsilon$, and given the relation $\chi = 4M^2 E^2 / L^2$, it follows that for a fixed angular momentum $L$, an increase in $\varepsilon$ necessitates a higher energy $E$. In the limit $\varepsilon \to 0$, we recover $\lim_{\varepsilon \to 0} \chi_*(\varepsilon) = \frac{4}{27}$. Combining this with $\chi = 4M^2 E^2 / L^2$, we confirm that for a Schwarzschild black hole pierced by a cosmic string, the energy condition for null geodesics imposed by $\Delta = 0$ reduces to the standard form: $E^2 = \frac{L^2}{27 M^2}.$ To further explore the effect of $\varepsilon$, we perform numerical integration and find that the integral $\int_{x_\text{init}}^{x_\text{f}} \frac{d x}{\sqrt{P(x)}}$ is a monotonically increasing function of $\varepsilon$ for a photon approaching a fixed radius from infinity (see Table I). This directly enhances the orbital winding behavior of the photon.
    
    \begin{center}
    	 \setlength{\tabcolsep}{10pt}
    	
    \textbf{Table I.} Integration results with $x_{\text{init}} = -1/3$ and $x_{\text{f}} = x_2 - 1 \times 10^{-4}$.
    
    \vspace{0.3em} 
    
    \begin{tabular}{cccccccc} 
    	\toprule
    	$\varepsilon$ & 0.1 & 0.25 & 0.4 & 0.55 & 0.7 & 0.85 & 1 \\
    	\midrule
    	$\displaystyle \int_{x_\text{init}}^{x_\text{f}} \frac{d x}{\sqrt{P(x)}}$
    	& 4.702 & 4.810 & 4.947 & 5.125 & 5.376 & 5.775 & 6.625 \\
    	\bottomrule
    \end{tabular}
    \end{center}
    
    \subsubsection*{(b) Orbits with constant $r$}
    
    \hspace{1.5em} Fig.5 presents three-dimensional plots of closed orbits at a fixed radius $r=r_0$, generated using the solution $\theta_-(\phi)$ from Eq.~(4.6). We now analyze the conditions for orbit closure and the corresponding periods. For a trajectory to close, the photon must return to its initial position, requiring the azimuthal angle to advance by $\delta\phi = 2\pi n$ ($n \in \mathbb{Z}$) while the polar angle $\theta$ repeats its cycle. Given the dependence of $\cos\theta$ on $\sin(\alpha\phi)$ in Eq.~(4.6), the periodicity condition $\sin[\alpha(\phi + 2n\pi)] = \sin(\alpha\phi)$ implies that $\alpha n = m$ ($m \in \mathbb{Z}$). If $\alpha$ is rational, written as a reduced fraction $\alpha = q_1/q_2$ (where $q_1, q_2$ are coprime integers), the smallest integer satisfying this condition is $n = q_2$. Consequently, the minimal azimuthal increment for orbit closure is $\delta\phi = 2\pi q_2$. In Fig.5, values of $\alpha$ are chosen such that the winding angles $\delta\phi$ correspond to $4\pi$, $6\pi$, and $8\pi$, respectively. The azimuthal increment $\delta\phi$ from case (a) aligns with the present analysis. Although radial variations in case (a) prevent the orbit from closing spatially, the angular motion retains its periodicity for rational $\alpha$. Given that $\theta$ and $\phi$ are coupled via Eq.~(3.8), this conclusion generalizes to all orbit types, provided the affine parameter $\eta$ spans a sufficiently large range. In contrast, when $\alpha$ is irrational, the orbit does not close; instead, it densely fills the latitudinal band bounded by the turning points in the polar angle $\theta$, as shown in Fig.6.
    \begin{figure}[H] 
    	\centering
    	
    	\begin{subfigure}[b]{0.3\textwidth}
    		\centering
    \includegraphics[width=\textwidth]{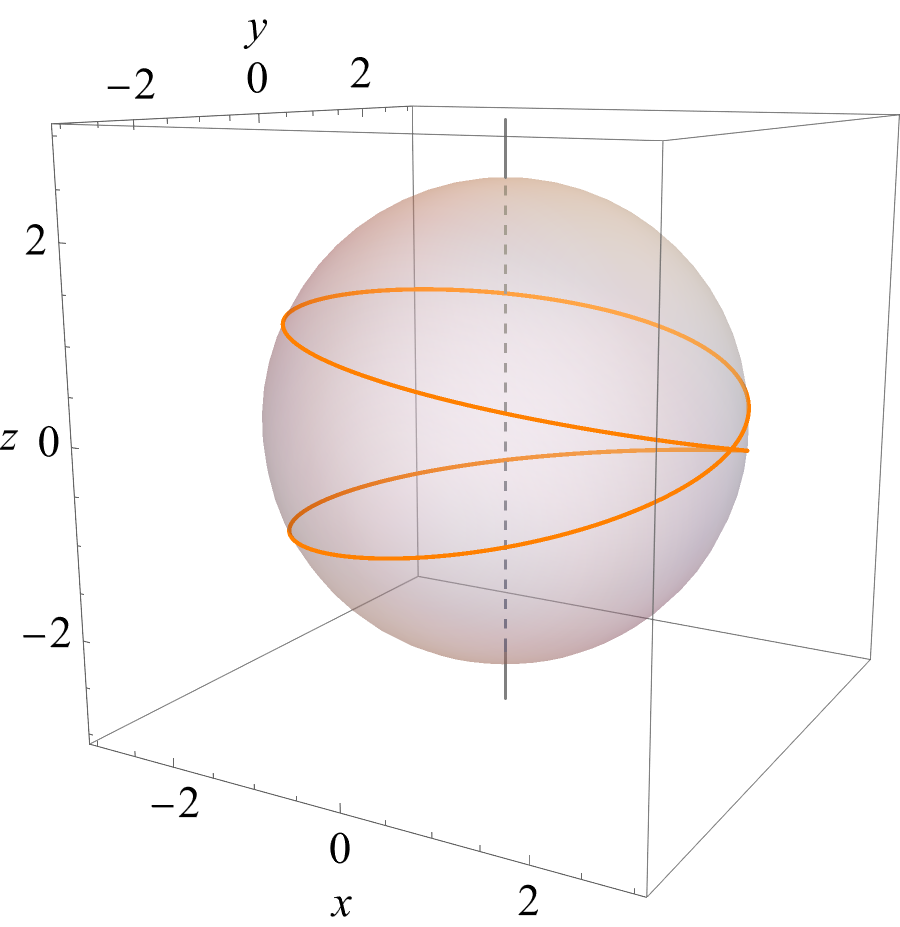}
    \caption{}
    \label{fig:a}
    \end{subfigure}
    \hfill 
    \begin{subfigure}[b]{0.3\textwidth}
    \centering
    \includegraphics[width=\textwidth]{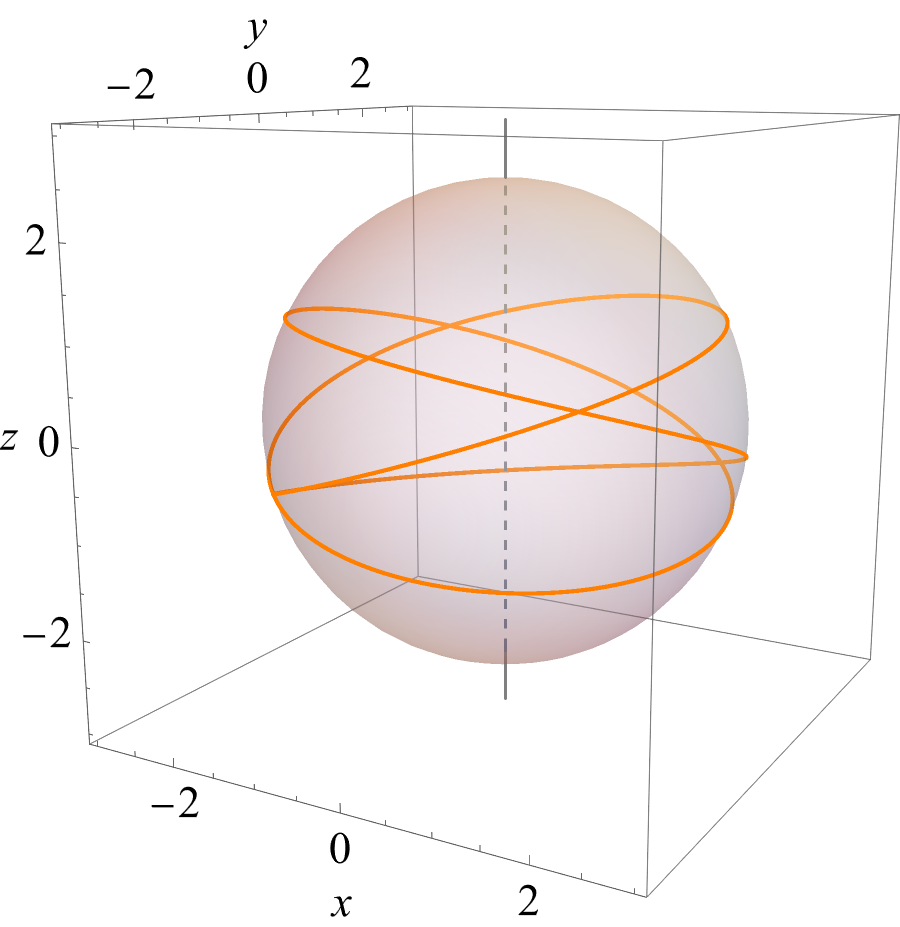}
    \caption{}
    \label{fig:b}
    \end{subfigure}
    \hfill 
    \begin{subfigure}[b]{0.3\textwidth}
    \centering
    \includegraphics[width=\textwidth]{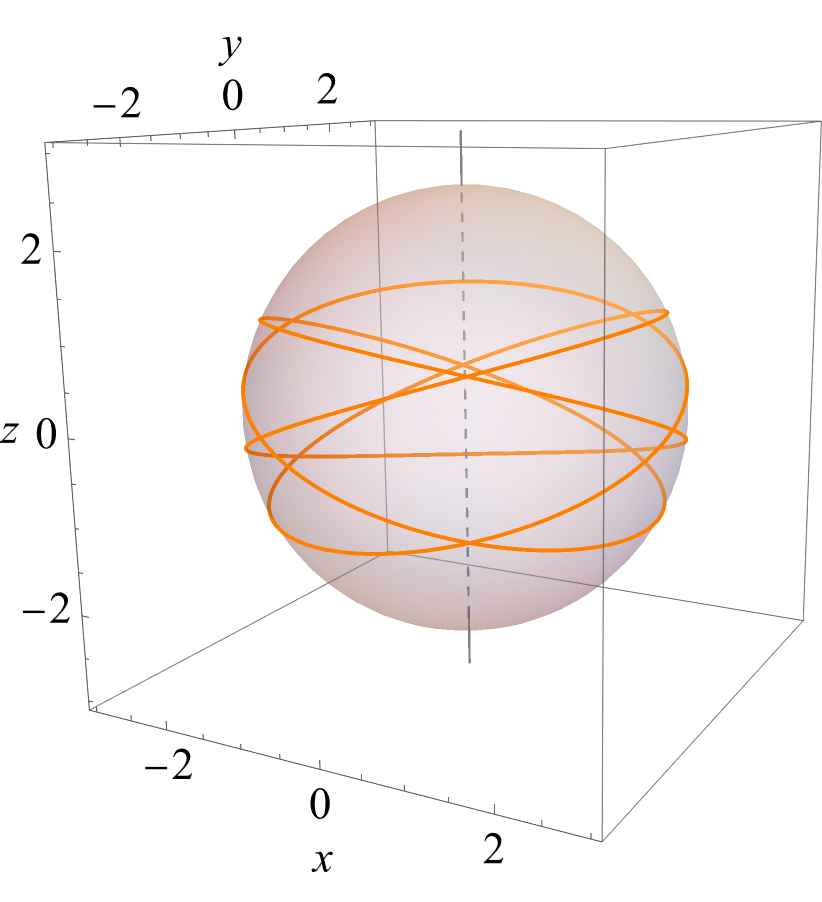}
    \caption{}
    \label{fig:c}
    \end{subfigure}
    
    \vspace{0.5cm} 
    
    \begin{subfigure}[b]{0.3\textwidth}
    \centering
    \includegraphics[width=\textwidth]{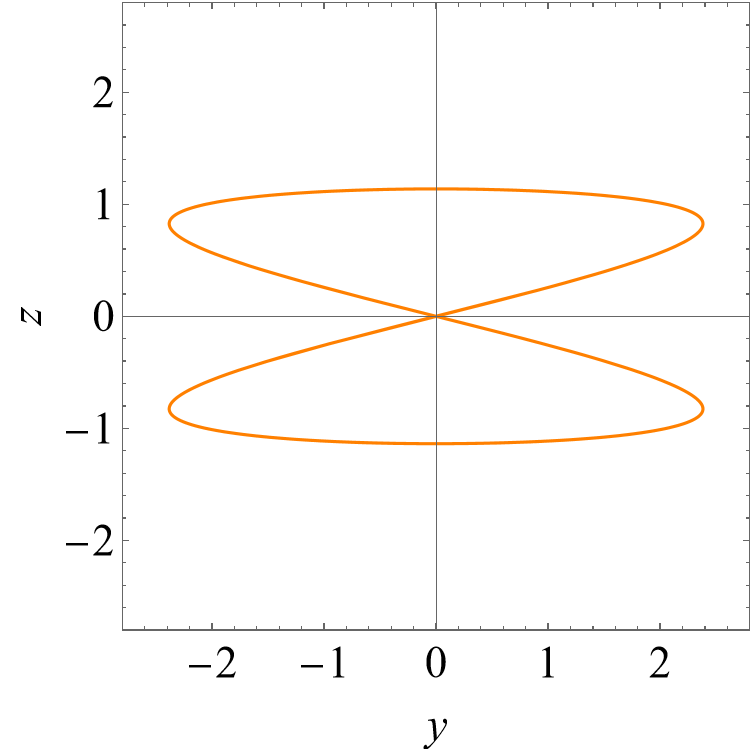}
    \caption{}
    \label{fig:d}
    \end{subfigure}
    \hfill
    \begin{subfigure}[b]{0.3\textwidth}
    \centering
    \includegraphics[width=\textwidth]{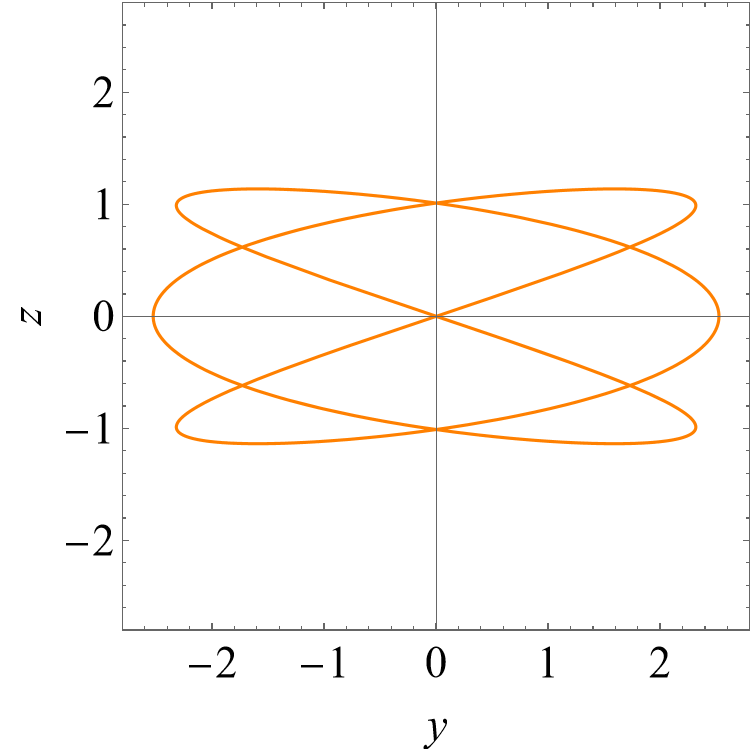}
    \caption{}
    \label{fig:e}
    \end{subfigure}
    \hfill
    \begin{subfigure}[b]{0.3\textwidth}
    \centering
    \includegraphics[width=\textwidth]{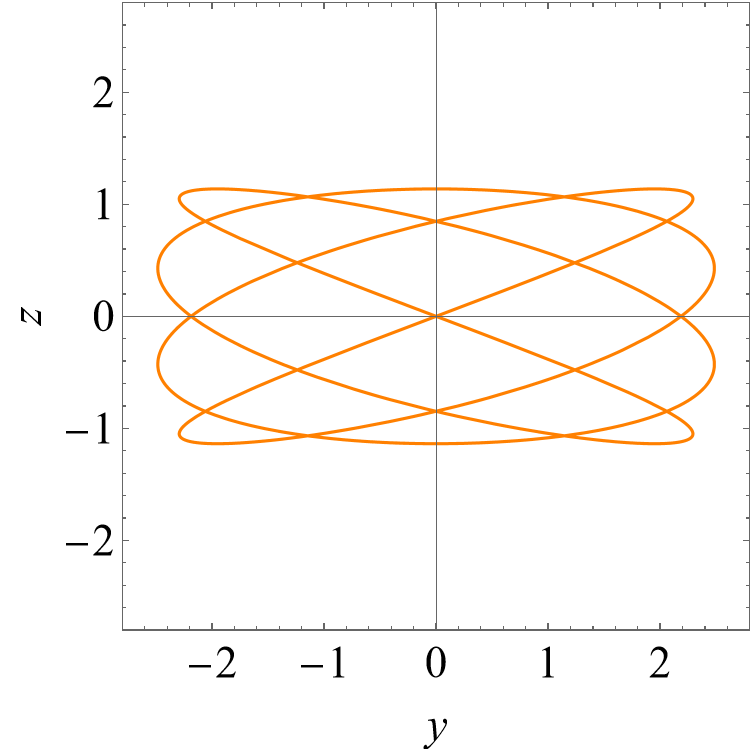}
    \caption{}
    \label{fig:f}
    \end{subfigure}
    
    \caption*{Fig.5: Periodic orbits of null geodesics at $r = r_0$, where (a,d), (b,e), and (c,f) correspond to $\alpha = 1/2$, $2/3$ and $3/4$, respectively. The parameters are chosen as $M = 1$, $\varepsilon = 0.6$, and $L/L_z = 1.12.$}
    \label{fig:total_3x2}
    \end{figure}
    
    \begin{figure}[H] 
    	\centering
    	
    	\begin{subfigure}[b]{0.3\textwidth}
    		\centering
    \includegraphics[width=\textwidth]{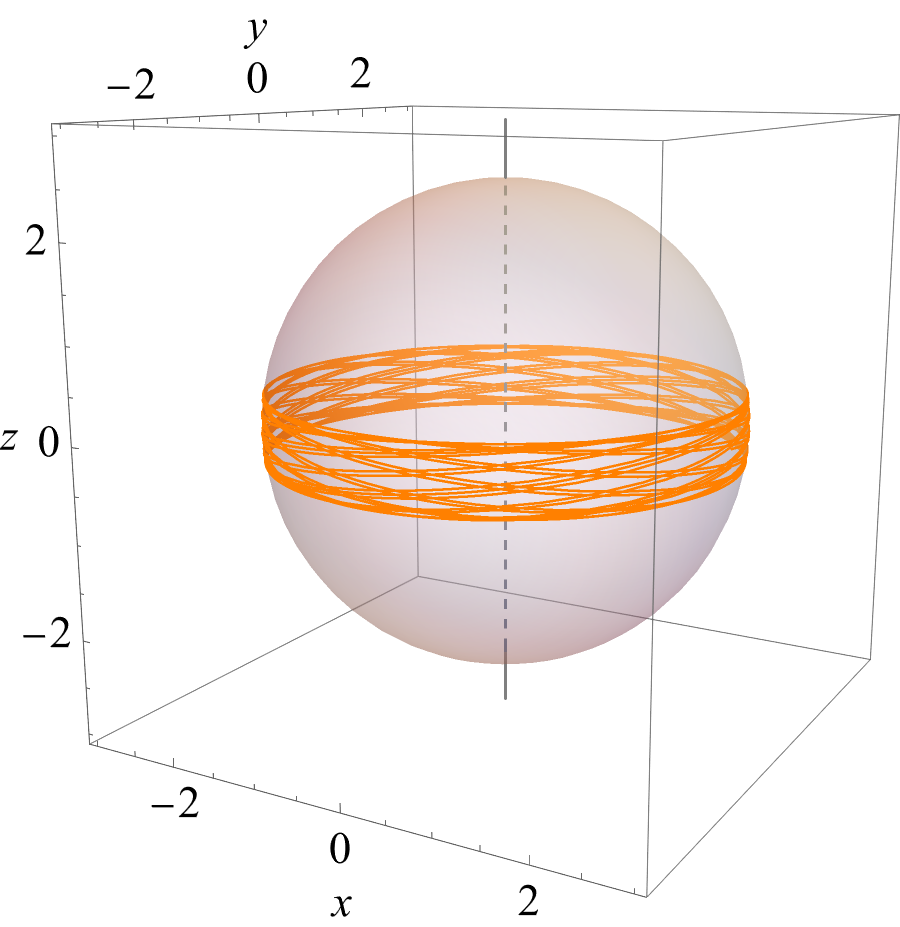}
    \caption{}
    \label{fig:a}
    \end{subfigure}
    \hfill 
    \begin{subfigure}[b]{0.3\textwidth}
    \centering
    \includegraphics[width=\textwidth]{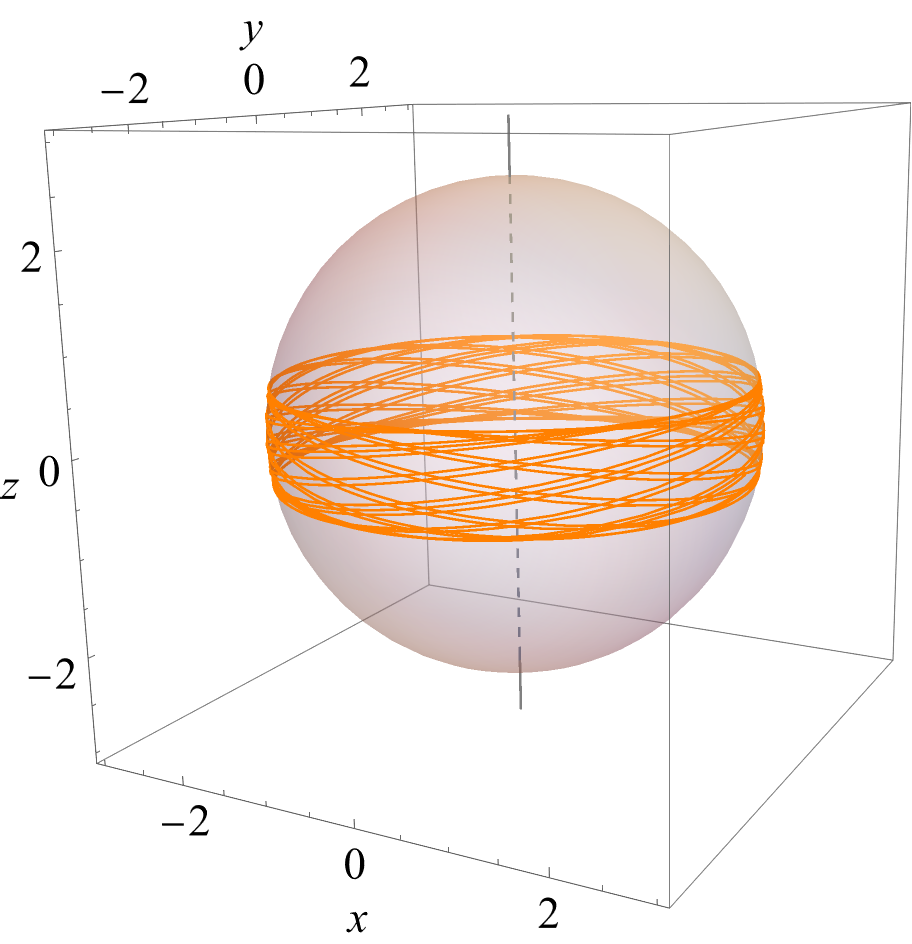}
    \caption{}
    \label{fig:b}
    \end{subfigure}
    \hfill 
    \begin{subfigure}[b]{0.3\textwidth}
    \centering
    \includegraphics[width=\textwidth]{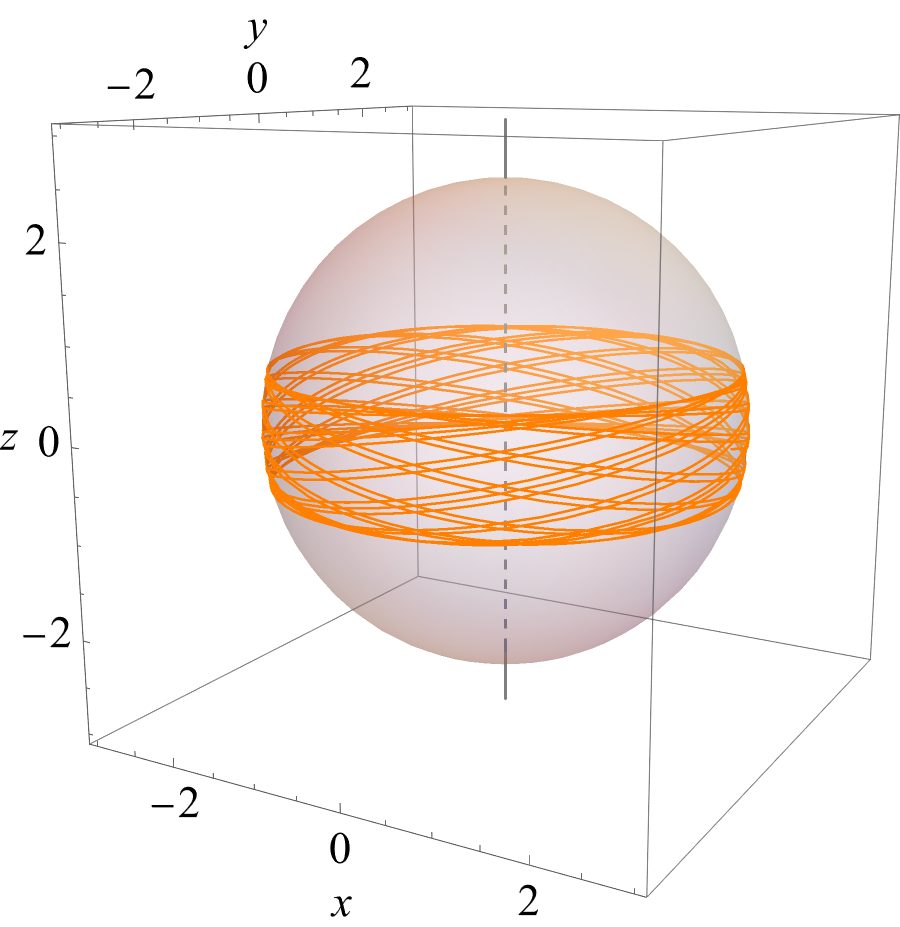}
    \caption{}
    \label{fig:c}
    \end{subfigure}
    \caption*{Fig.6: Non-closed orbits of null geodesics at $r = r_0$, with (a), (b), and (c) corresponding to $L/L_z =1.01$, $1.02$ and $1.03$, respectively. The parameters are chosen as $M = 1$, $\varepsilon = 0.6$, and $\alpha = \sqrt{2/3}.$}
    \label{fig:total_3x2}
\end{figure}
    
    \subsubsection*{(c) Orbits entering the horizon}
    
    \hspace{1.5em}Infinitesimal perturbations to null geodesics on the constant-radius orbit $r_0$ can lead them to either spiral inward into the black hole or escape to infinity. Here, we focus on the ingoing trajectories, as illustrated in Fig.6. A comparison of panels (a) and (c) reveals that decreasing $\alpha$ increases the number of windings prior to encountering the singular potential barrier. This spiraling behavior, primarily localized in the vicinity of $r_0$, aligns with the analysis of case (a): the radial coordinate $r$ evolves slowly with respect to the affine parameter $\eta$ in this region, causing the geodesic to linger near $r_0$ and execute multiple loops. Similarly, numerical integration results indicate that an increase in the charge parameter $\varepsilon$ enhances the integral value (Table II), thereby intensifying the winding behavior. 
    
    \begin{figure}[H]
    	\centering
    	
    	\begin{subfigure}[b]{0.48\textwidth}
    		\centering
    		\includegraphics[width=\textwidth]{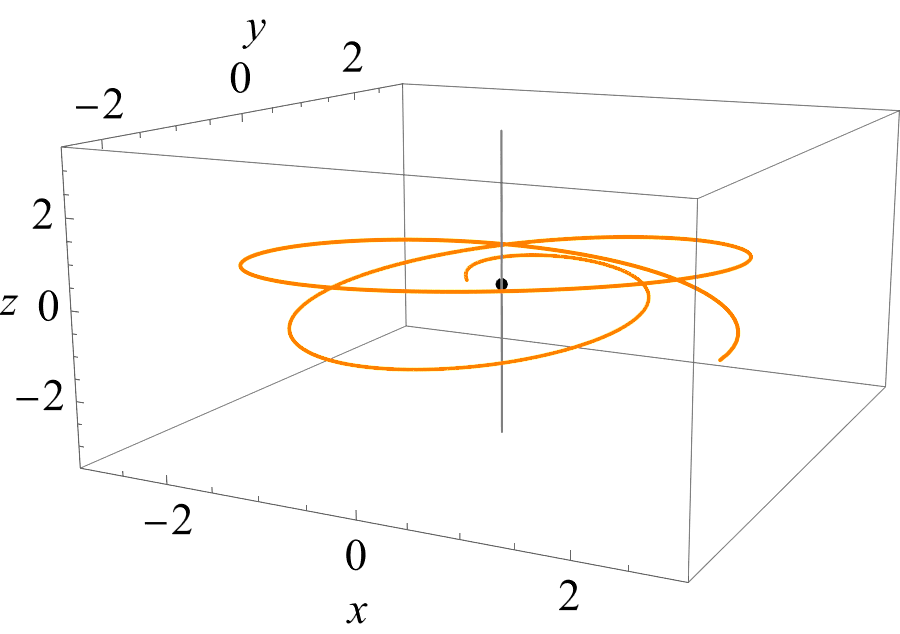} 
    		\caption{}
    		\label{fig:1a}
    	\end{subfigure}
    	\hfill 
    	\begin{subfigure}[b]{0.48\textwidth}
    		\centering
    		\includegraphics[width=\textwidth]{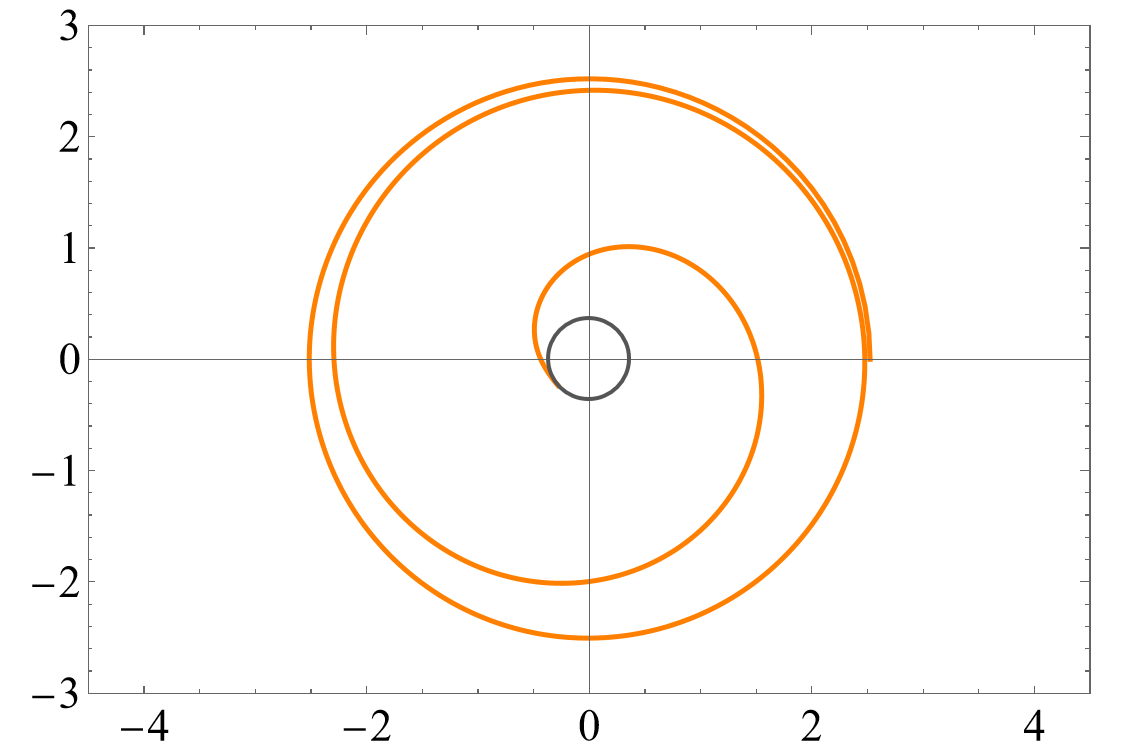}
    		\caption{}
    		\label{fig:1b}
    	\end{subfigure}
    	
    	\vspace{0.5cm} 
    	
    	\begin{subfigure}[b]{0.48\textwidth}
    		\centering
    		\includegraphics[width=\textwidth]{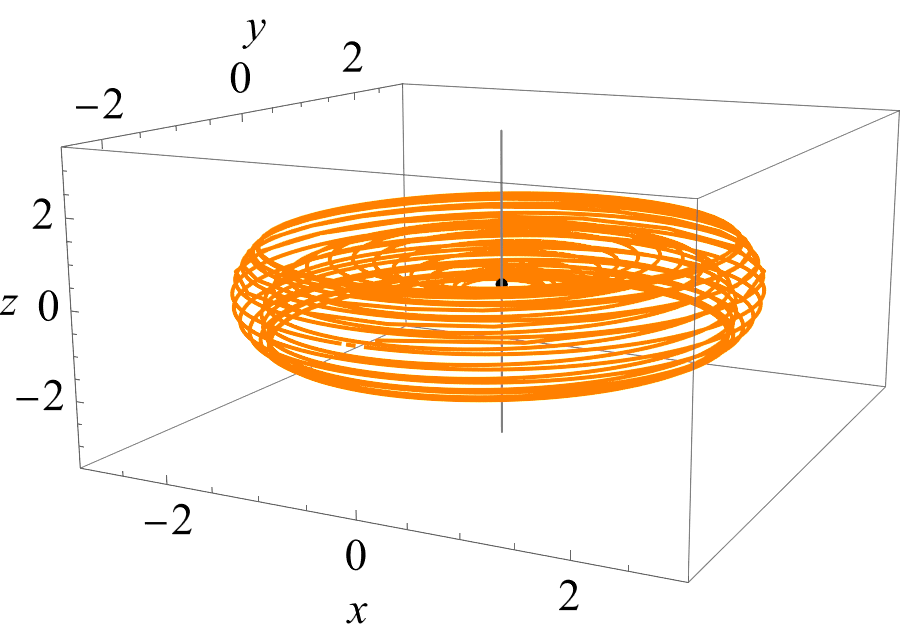}
    		\caption{}
    		\label{fig:1c}
    	\end{subfigure}
    	\hfill
    	\begin{subfigure}[b]{0.48\textwidth}
    		\centering
    		\includegraphics[width=\textwidth]{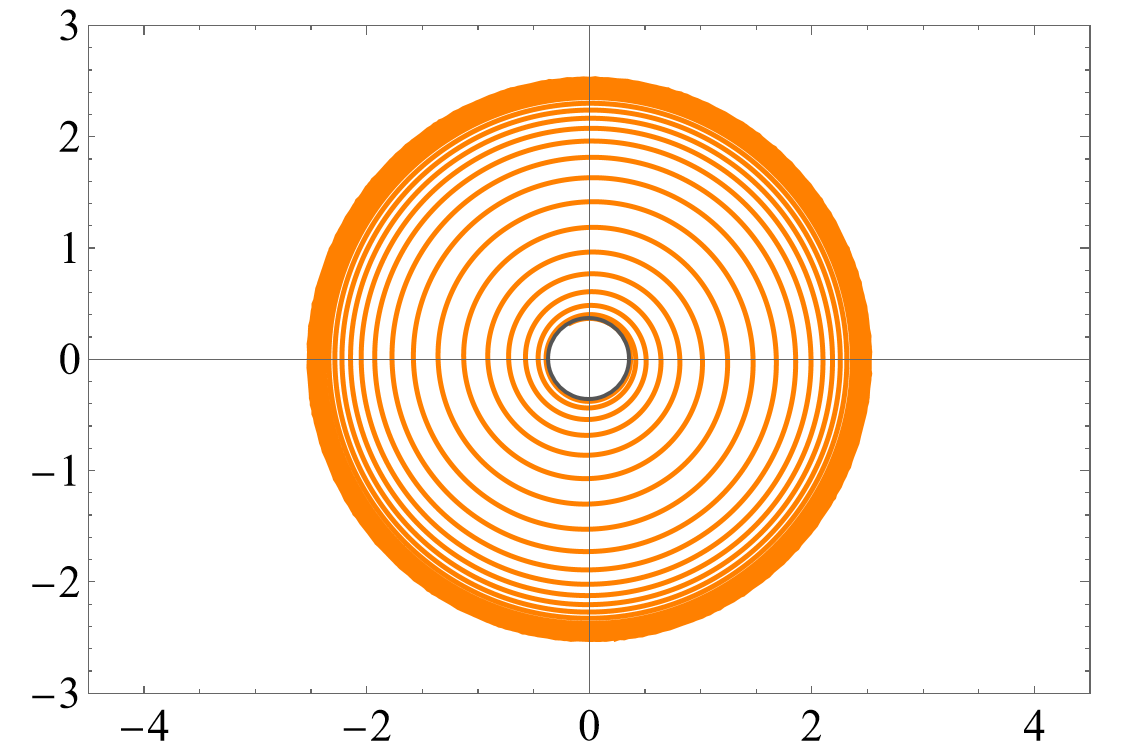}
    		\caption{}
    		\label{fig:1d}
    	\end{subfigure}
    	\vspace{0.0cm}
    	\caption*{Fig.7: Light rays starting from a radius slightly below $r_0$, falling into the black hole and encountering an  infinite potential barrier ($\Delta = 0$). (a,b) $\alpha = 0.6$; (c,d) $\alpha = 0.05$. $M=1$, $L/L_z = 1.12$, and $\varepsilon = 0.6$ are used throughout the figure. Panels (b,d) show $\phi$-$r$ trajectories in polar coordinates, with the gray ring marking the singularity barrier.}
    	\label{fig:total}
    \end{figure}

     \begin{center}
    	\setlength{\tabcolsep}{10pt}
    	
    	\textbf{Table II.} Integration results with $x_{\text{init}} = x_2 + 1 \times 10^{-4}$ and $x_{\text{f}} = x_3$.
    	
    	\vspace{0.3em} 
    	
    	\begin{tabular}{cccccccc} 
    		\toprule
    		$\varepsilon$ & 0.1 & 0.25 & 0.4 & 0.55 & 0.7 & 0.85 & 1 \\
    		\midrule
    		$\displaystyle \int_{x_\text{init}}^{x_\text{f}} \frac{d x}{\sqrt{P(x)}}$
    		& 5.358 & 5.463 & 5.595 & 5.769 & 6.014 & 6.406 & 7.248 \\
    		\bottomrule
    	\end{tabular}
    \end{center}
     
    \subsection*{5.2 The case $\Delta < 0$}
    
    \hspace{1.5em}This regime corresponds to $\chi > \chi_*$, in which ingoing null geodesics possess sufficient energy to surmount the critical effective potential at $r_0$ and therefore cross both horizons before encountering the singular potential barrier. Fig.8 depicts this scenario, illustrating trajectories incident from infinity. Notably, decreasing $\alpha$ results in a higher frequency of orbital revolutions before reaching the barrier. Numerical integration confirms that for a fixed $\chi$ (i.e., constant $E/L$), increasing $\varepsilon$ similarly enlarges the value of the integral $\int_{x_{\text{init}}}^{x_f} \frac{dx}{\sqrt{P(x)}},$ implying that a larger charge $Q$ yields more pronounced winding (Table III). This trend can be qualitatively understood from the equation of motion (3.4). For a fixed $E/L$, an increase in $\varepsilon$ leads to a higher effective potential, which causes the radial coordinate $r$ to evolve more slowly with respect to the affine parameter $\eta$. Since the rate of change of the angular variables with respect to $\eta$ remains unaffected, a given radial displacement $\delta r$ corresponds to a larger azimuthal shift $\delta\phi$. This explains the numerical results in Table III.

    \begin{figure}[htbp]
    	\centering
    	
    	\begin{subfigure}[b]{0.48\textwidth}
    		\centering
    		\includegraphics[width=\textwidth]{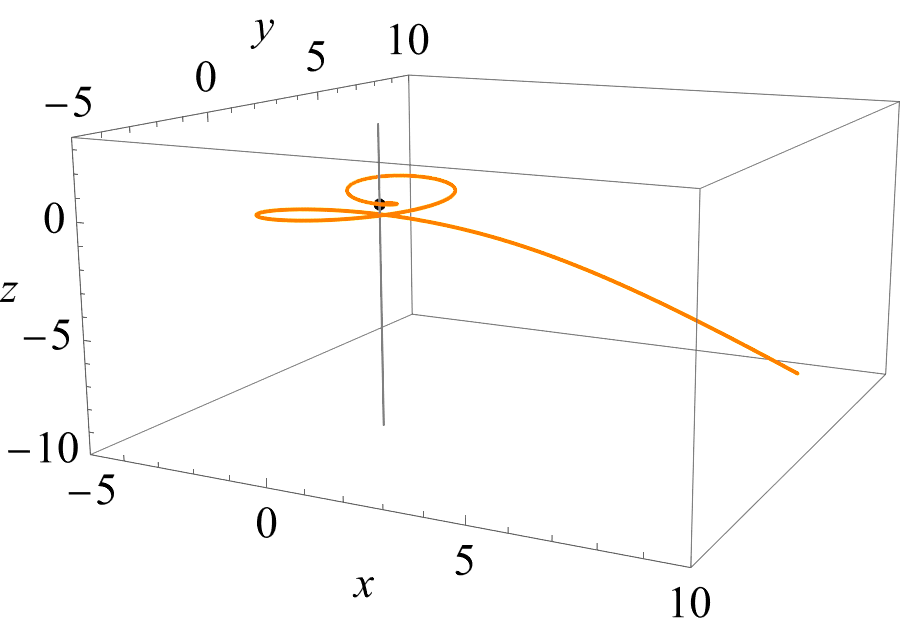} 
    		\caption{}
    		\label{fig:1a}
    	\end{subfigure}
    	\hfill 
    	\begin{subfigure}[b]{0.48\textwidth}
    		\centering
    		\includegraphics[width=\textwidth]{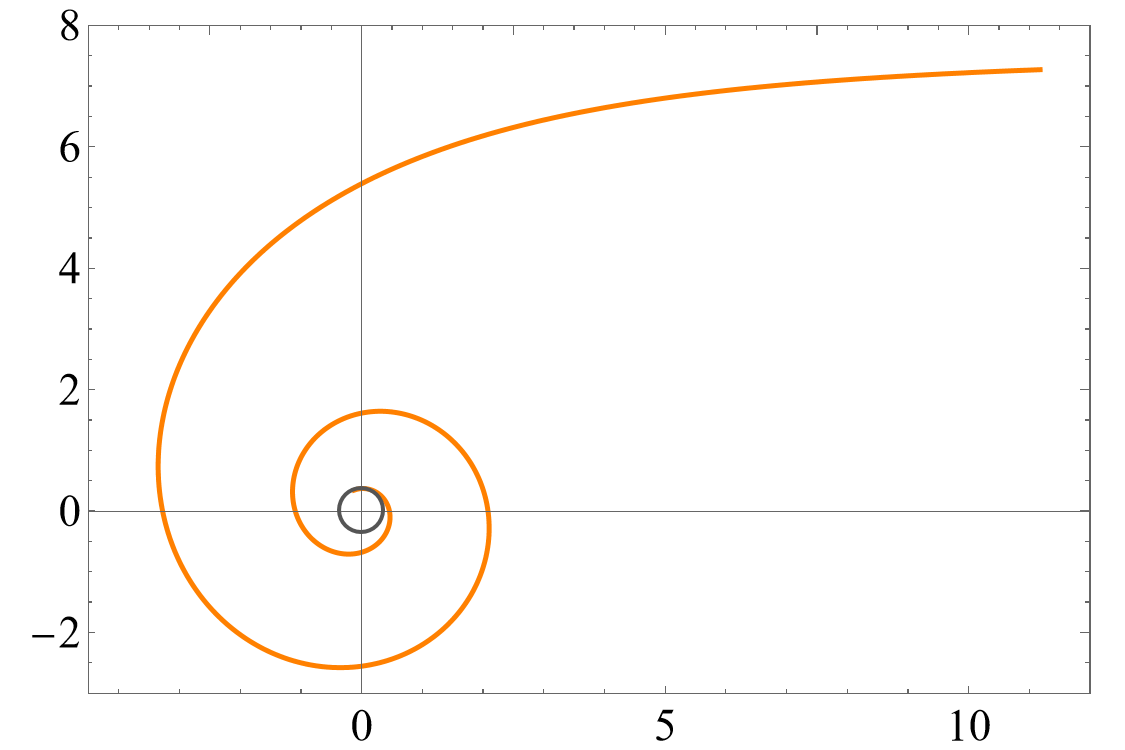}
    		\caption{}
    		\label{fig:1b}
    	\end{subfigure}
    	
    	\vspace{0.5cm} 
    	
    	\begin{subfigure}[b]{0.48\textwidth}
    		\centering
    		\includegraphics[width=\textwidth]{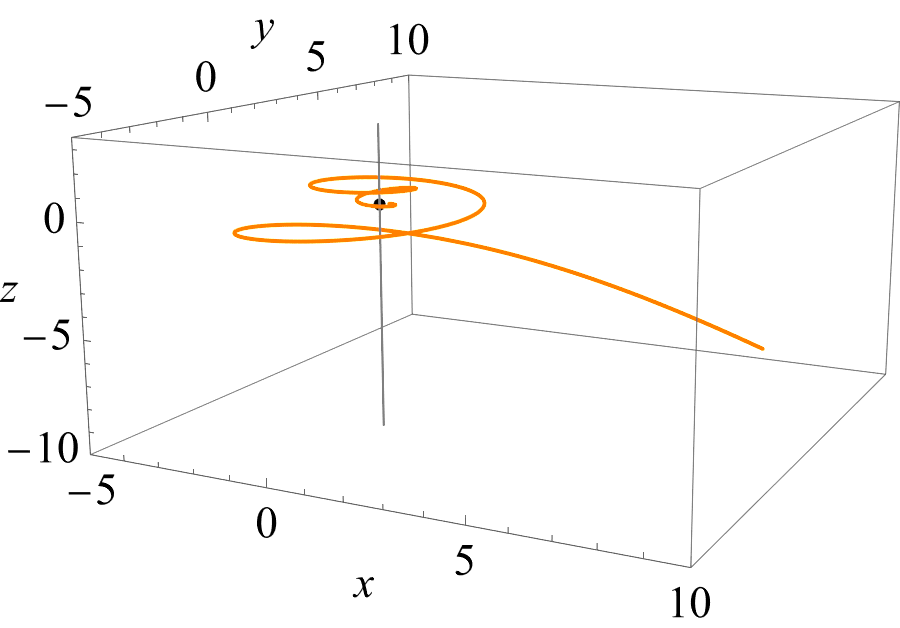}
    		\caption{}
    		\label{fig:1c}
    	\end{subfigure}
    	\hfill
    	\begin{subfigure}[b]{0.48\textwidth}
    		\centering
    		\includegraphics[width=\textwidth]{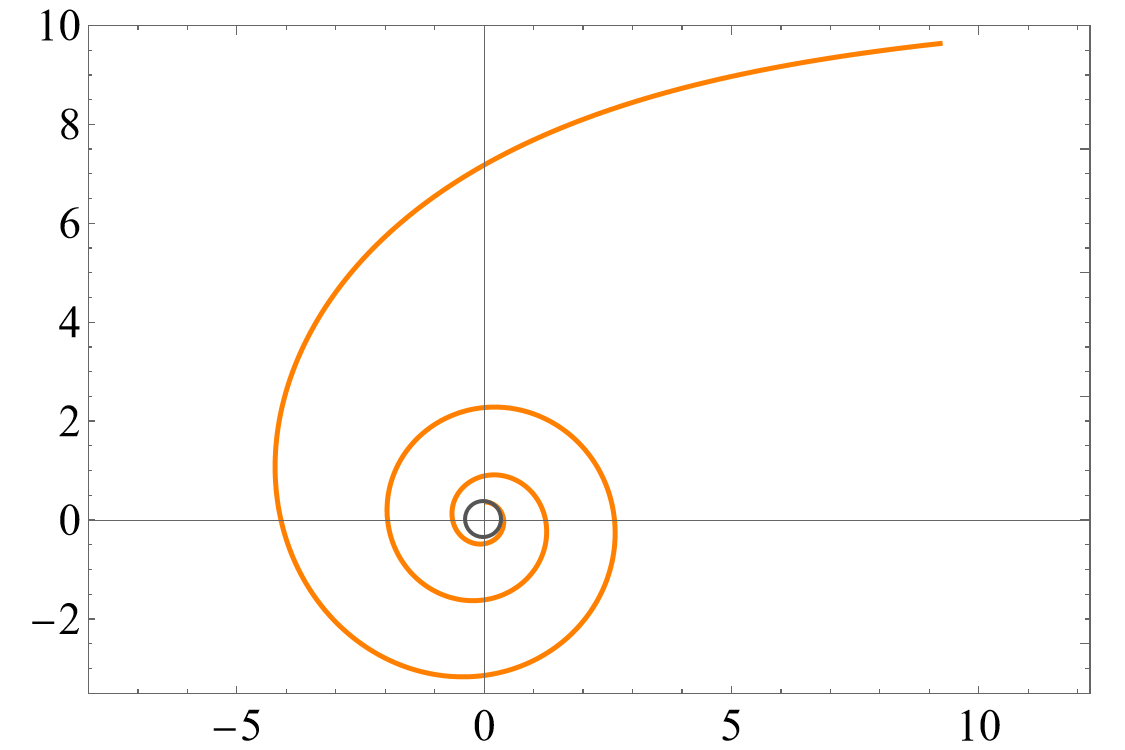}
    		\caption{}
    		\label{fig:1d}
    	\end{subfigure}
    	\vspace{0.0cm}
    	\caption*{Fig.8: Simulation results for $\Delta < 0$ with (a,c) $\alpha = 0.35$ and (b,d) $\alpha = 0.25$. Other parameters: $M=1$, $L/L_z = 1.3$, and $\varepsilon = 0.6$.}
    	\label{fig:total}
    \end{figure}
    
     \begin{center}
    	\setlength{\tabcolsep}{10pt}
    	
    	\textbf{Table III.} Integration results from $x_{\mathrm{init}} = -1/3$ to $x_{\mathrm{f}} = x_2$ for $\chi = 0.5$.
    	
    	\vspace{0.3em} 
    	
    	\begin{tabular}{cccccccc} 
    		\toprule
    		$\varepsilon$ & 0.1 & 0.25 & 0.4 & 0.55 & 0.7 & 0.85 & 1 \\
    		\midrule
    		$\displaystyle \int_{x_\text{init}}^{x_\text{f}} \frac{d x}{\sqrt{P(x)}}$
    		& 1.998 & 2.031 & 2.070 & 2.114 & 2.166 & 2.223 & 2.271 \\
    		\bottomrule
    	\end{tabular}
    \end{center}

    \subsection*{5.3 The case $\Delta > 0$}
    
    \hspace{1.5em}In this scenario, the equation $P(x) = 0$ yields four distinct real roots, ordered as $x_1 < -1/3 < x_2 < x_3 < x_4$. We focus on null geodesics that originate from spatial infinity, approach the black hole, are reflected by the critical effective potential, and subsequently escape back to infinity---a process governed by Eq.~(4.10). Correspondingly, the relevant integration interval is restricted to $[-1/3, x_2]$.
    
    Fig.9 illustrates the impact of varying $\alpha$ on these bounce orbits. As anticipated, decreasing $\alpha$ increases the number of loops light rays execute around the black hole. Simulations further reveal that for a light ray originating from spatial infinity with initial polar angle $\theta_0 = \pi - \arcsin(L_z/L)$, the final polar angle remains invariant under changes in $\alpha$. For the specific parameters in Fig.9, the escaping light ray asymptotically approaches a fixed polar angle $\theta_{\infty} \approx 0.980$. In contrast, the azimuthal angle $\phi$---closely tied to the winding behavior---is significantly affected by $\alpha$. Radial motion is dictated by the polynomial $P(x)$, which is independent of $\alpha$. Consequently, the integral $\int \frac{dx}{\sqrt{P(x)}}$ along the path $-1/3 \to x_2 \to -1/3$ is determined solely by $\varepsilon$ and $\chi$ (see Eq.~4.6). This accounts for the numerical observation that radial evolution remains unchanged when $\alpha$ varies. Conversely, from the relation $\cot^2 \theta = (L^2/L_z^2 - 1) \sin^2(\alpha \phi)$, for a fixed change in $\theta$, the final value of $\phi$ explicitly depends on $\alpha$. Thus, $\alpha$ affects only the accumulated azimuthal rotation. In addition to the effect of $\alpha$, we examine the influence of the charge parameter $\varepsilon$ at fixed $\chi$. The radial integral values in Table IV show that increasing $\varepsilon$ reduces the integral, implying that the winding behavior is weakened. Moreover, unlike $\alpha$, varying $\varepsilon$ alters the polar angle $\theta$ of geodesics that eventually escape to infinity, since $\varepsilon$ appears explicitly in $P(x)$.
    
    The spacetime of a Reissner–Nordström black hole is perfectly spherical. Consequently, photons reflected by the effective potential remain confined to a fixed orbital plane and exhibit no three-dimensional winding. In contrast, the cosmic string breaks spherical symmetry, causing null geodesics to evolve in $z$ and leading to non‑planar orbits. Fig.10 displays typical photon orbits for a significant cosmic string parameter $\alpha$: for an initial polar angle $\theta_0 = \pi - \arcsin(L_z/L)$, the orbit assumes a spindle-like shape; for orbits incident from infinity perpendicular to the cosmic string, they form a semi-spindle structure. These features differ markedly from the planar orbits and could serve as distinctive observational signatures.
    
    We now briefly discuss the special case $L/L_z = 1$, where null geodesics are confined to the equatorial plane. For null geodesics that originate from spatial infinity, approach the black hole, and are then reflected back to infinity, we examine the deflection angle in this scenario. At the radius of closest approach $r_b$ (corresponding to $x_b$), we have $t(x_b) = e_1$. Consequently, the elliptic integral $F(\varphi, k)$ reduces to the complete elliptic integral of the first kind $F(\pi/2, k) = K(k)$. Substituting the radial integral result Eq.~(4.20) into Eq.~(4.2) and noting that the initial and final positions of the photon correspond to $x = -1/3$, we obtain
    \begin{equation}
    	\begin{split}
    		\frac{2}{\sqrt{e_1 - e_3}} \Bigg[ K(k) - F\bigg( \arcsin &\sqrt{\frac{e_1 - e_3}{t(-1/3) - e_3}}, k \bigg) \Bigg] \\
    		&= \frac{1}{2} \left\{ \alpha\,\delta\phi + \arctan\!\left[ \frac{(L/L_z - 1) \sin(2\alpha\,\delta\phi)}{(L/L_z + 1) - (L/L_z - 1) \cos(2\alpha\,\delta\phi)} \right] \right\}.
    	\end{split}\tag{5.3}
    \end{equation}
    Setting $L/L_z=1$, we directly obtain the total change in the azimuthal angle along the null geodesic as
    \begin{equation}
    	\delta\phi = \frac{4}{\alpha \sqrt{e_1 - e_3}} \left[ K(k) - F\left( \arcsin \sqrt{\frac{e_1 - e_3}{t(-1/3) - e_3}}, k \right) \right].\tag{5.4}
    \end{equation}
    In the weak‑field limit, where $E^2/V_{\text{eff}}^c \to 0^+$, the deflection angle is defined as $\Delta\phi = \delta\phi - \pi$. When $\alpha = 1$, this expression recovers the deflection angle for light grazing an RN black hole. An increase in the string energy density enhances the deflection angle of light. As $\alpha \to 0$, photons may orbit the black hole-string system multiple times, causing the classical deflection angle to break down. In this regime, to maintain a meaningful definition, it is  appropriate to adopt the modulated deflection angle $\Delta\phi_{\text{mod}} = \delta\phi - (2n+1)\pi$, with $n \in \mathbb{Z}$ chosen such that $-\pi < \Delta\phi_{\text{mod}} \leq \pi$. While this definition remains valid for strong gravitational fields ($E^2/V_{\text{eff}}^c \to 1^-$), our analysis here focuses on light deflection induced by classical weak gravitational effects. Fig.11 clearly demonstrates that for a fixed $\chi$, an increase in charge suppresses $\delta\phi$, whereas an increase in $\chi$ enhances it. The enhancing effect of $\chi$ can be qualitatively elucidated through the equation of motion (3.4). An increase in $\chi$ corresponds to an increase in the ratio $E/L$. For a qualitative analysis, assuming fixed $\varepsilon$ and $L$ leaves the effective potential unchanged. Consequently, a larger $E$ causes the null geodesics to encounter the potential barrier at a smaller turning radius $r_b$, probing a region of stronger gravity and thus yielding a larger deflection angle. The suppression of the deflection angle by the charge is evident from the integral results in Table IV: increasing $\varepsilon$ reduces the radial integral, which in turn gives a smaller $\Delta\phi$. To facilitate a direct comparison with light deflection in RN black hole geometry, we can consistently express our parameter $\chi$ as $\chi = 4M^2/b_{\text{im}}^2$, with $b_{\text{im}} = L/E$ being the conventional impact parameter used in gravitational lensing studies. Ref. [31] points out that both an increase in the charge $Q$ and an increase in $b_{\text{im}}$ lead to a reduction of the deflection angle in the RN spacetime, in full agreement with our results.
    
    \begin{figure}[H]
    	\centering
    	
    	\begin{subfigure}[b]{0.48\textwidth}
    		\centering
    		\includegraphics[width=\textwidth]{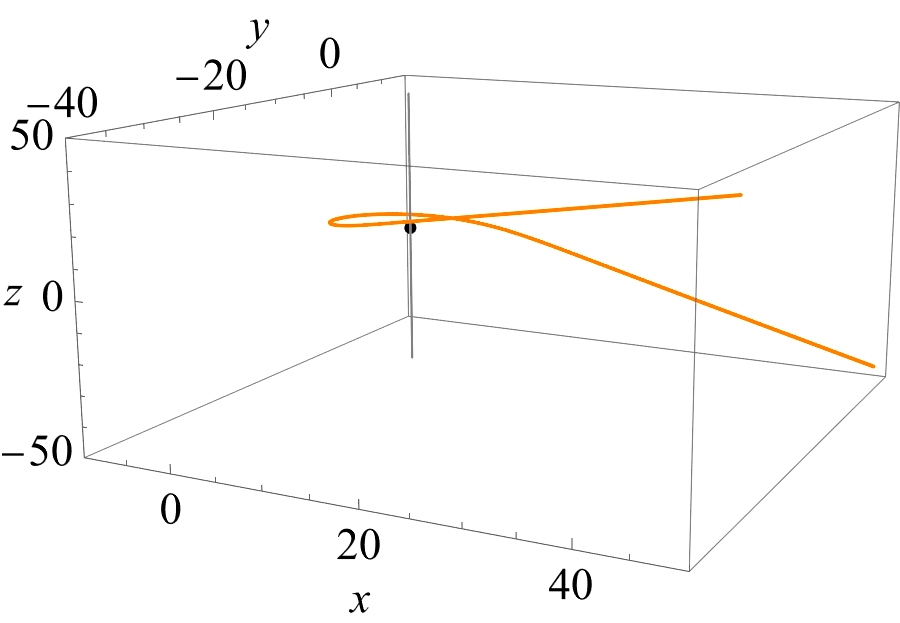} 
    		\caption{}
    		\label{fig:1a}
    	\end{subfigure}
    	\hfill 
    	\begin{subfigure}[b]{0.48\textwidth}
    		\centering
    		\includegraphics[width=\textwidth]{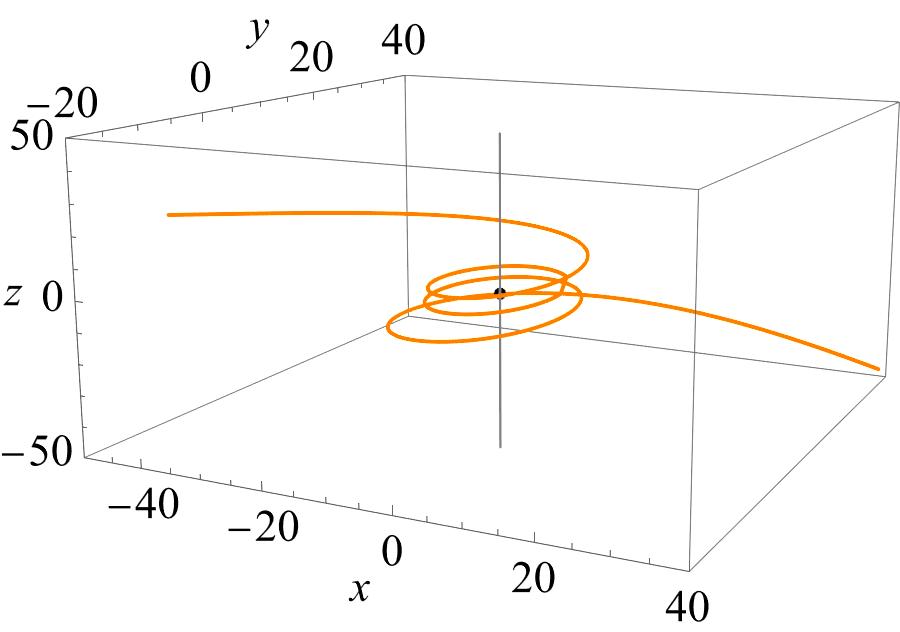}
    		\caption{}
    		\label{fig:1b}
    	\end{subfigure}
    	\vspace{0.0cm}
    	\caption*{Fig.9: Null geodesics from infinity reflected by the effective potential for (a) $\alpha = 0.6$ and (b) $\alpha = 0.15$. The parameters are $L/L_z=1.3$ and $\varepsilon = 0.6$.}
    	\label{fig:total}
    \end{figure}
    
    \begin{figure}[H]
    	\centering
    	
    	\begin{subfigure}[b]{0.48\textwidth}
    		\centering
    		\includegraphics[width=\textwidth]{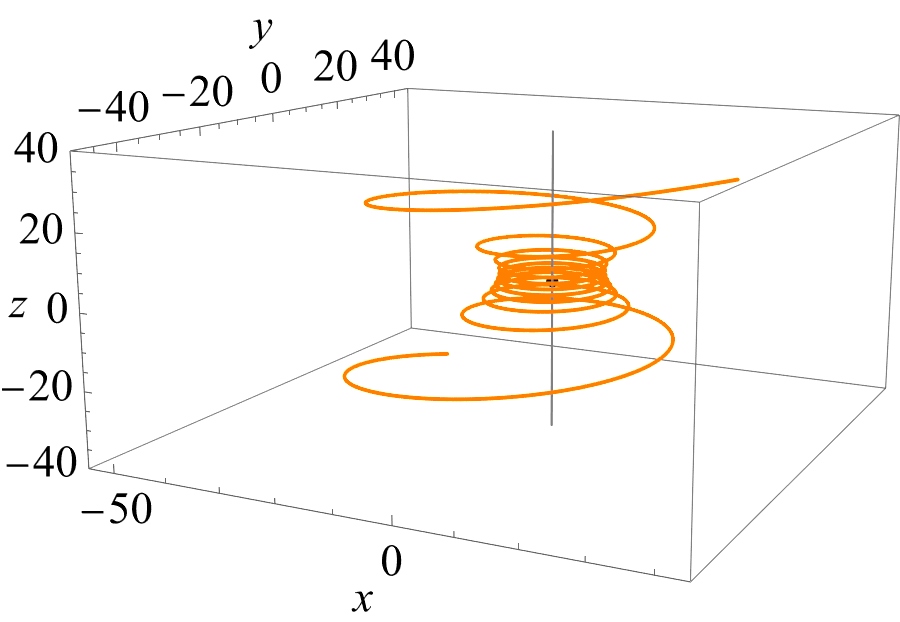} 
    		\caption{}
    		\label{fig:1a}
    	\end{subfigure}
    	\hfill 
    	\begin{subfigure}[b]{0.48\textwidth}
    		\centering
    		\includegraphics[width=\textwidth]{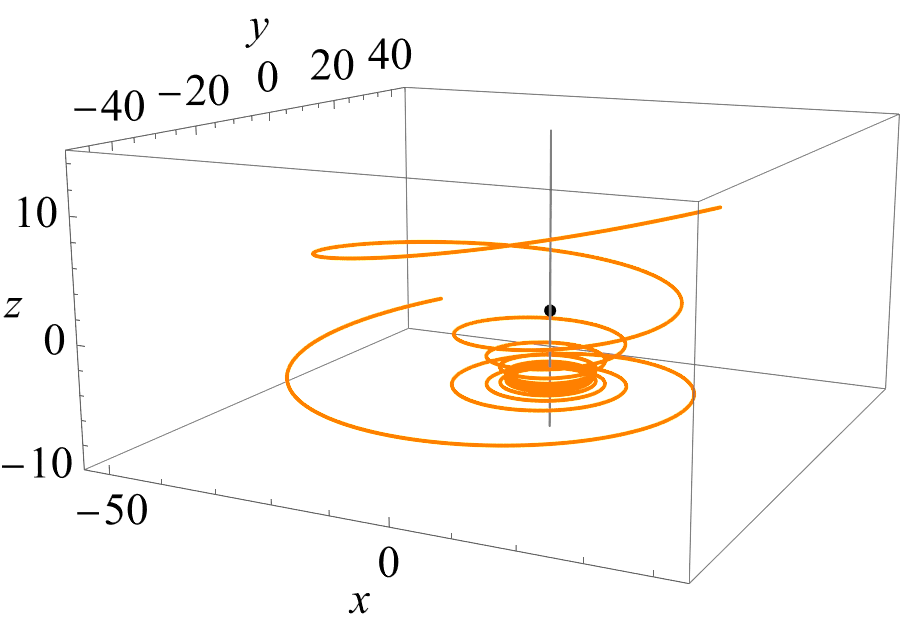}
    		\caption{}
    		\label{fig:1b}
    	\end{subfigure}
    	\vspace{0.0cm}
    	\caption*{Fig.10: Orbital configurations with significant cosmic string effect ($\alpha = 0.05$): (a) initial polar angle $\theta_0 = \pi - \arcsin(L_z/L)$; (b) perpendicular incidence from infinity. Other parameters as in Fig.8.}
    	\label{fig:total}
    \end{figure}
    
    \begin{figure}[H]
     	\centering
     	\includegraphics[width=0.53\textwidth]{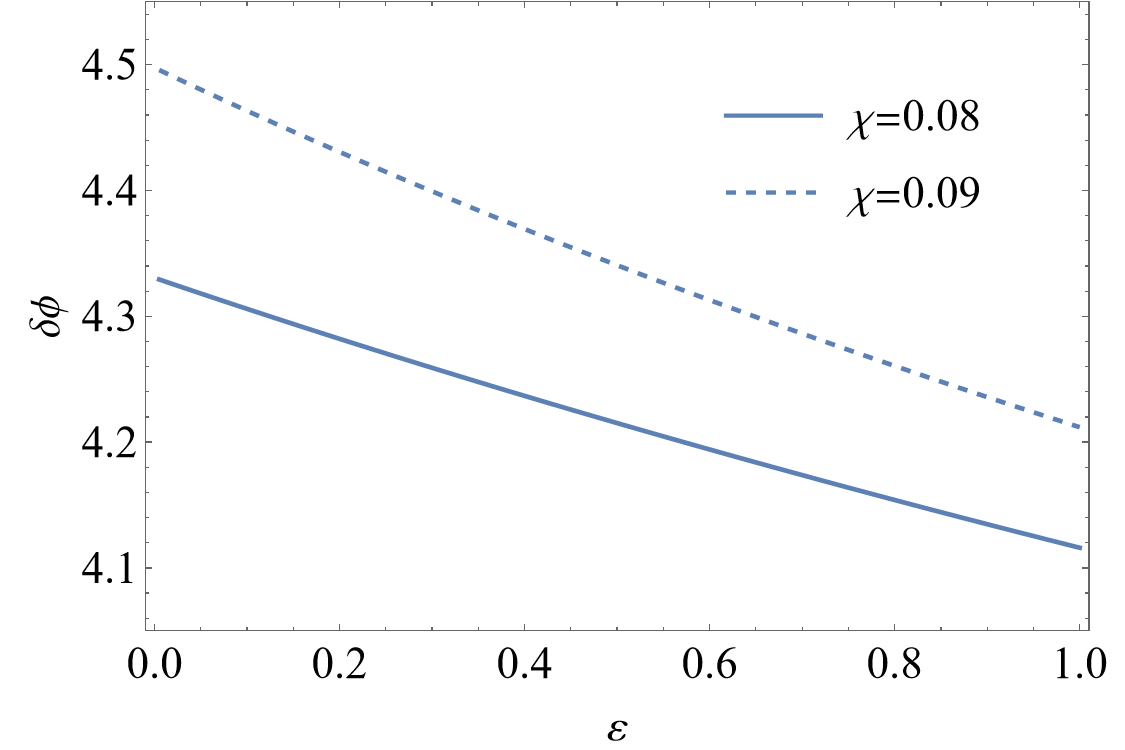}  
     	\caption*{Fig.11: Variation of $\delta\phi$ with $\varepsilon$ on the equatorial plane with $\alpha=0.98$.}
     	\label{fig:single}
     \end{figure}

      \begin{table}[H]  
     	\centering
     	\caption*{\textbf{Table IV.} Integration results along the path $x_{\mathrm{init}} = -1/3 \to x_b \to -1/3$ for $\chi = 0.08$.}
     	\label{tab:integral_results}  
     	\setlength{\tabcolsep}{10pt}
     	\begin{tabular}{cccccccc}
     		\toprule
     		$\varepsilon$ & 0.1 & 0.25 & 0.4 & 0.55 & 0.7 & 0.85 & 1 \\
     		\midrule
     		$\displaystyle 2\int_{x_\text{init}}^{x_b} \frac{d x}{\sqrt{P(x)}}$
     		& 2.110 & 2.093 & 2.076 & 2.060 & 2.045 & 2.031 & 2.017 \\
     		\bottomrule
     	\end{tabular}
     \end{table}

     \section*{6. Conclusions}
     
     In this work, we investigate null geodesics in the spacetime of an RN black hole pierced by a cosmic string. Based on the radial equation of motion, we classify the geodesics into three regimes depending on whether the squared energy $E^2$ exceeds, equals, or falls below the critical effective potential $V_{\mathrm{eff}}^c$. This classification can be equivalently derived from the analysis of the polynomial $P(x)$, and is entirely governed by the critical parameter $\chi_*$. Furthermore, by analyzing the root distribution of the quartic polynomial $P(x)$ within each regime, we obtain analytical solutions in two equivalent forms: one expressed via the inverse Weierstrass elliptic function, and the other via elliptic integrals. Our analysis of the angular motion reveals that photons are strictly confined to the equatorial plane if and only if the total angular momentum $L$ equals its axial component $L_z$ along the cosmic string. Otherwise, the trajectories are non-planar, representing a striking departure from the behavior of spherically symmetric black holes.
     
     Building upon these analytical solutions, we subsequently employ numerical simulations to trace the propagation trajectories of null geodesics across the different regimes. Our results reveal that the cosmic string parameter $\alpha$ universally enhances the winding behavior of null geodesics in all three regimes, whereas the effect of the black hole charge $Q$ is regime-dependent. Interestingly, when $\alpha$ is a reduced fraction $\alpha = q_1/q_2$ (with coprime integers $q_1$ and $q_2$), circular orbits at $r = r_0$ become strictly periodic, exhibiting an azimuthal increment of $\delta\phi = 2\pi q_2$ per period. Furthermore, in scenarios involving reflection off the potential barrier, a sufficiently large cosmic string energy density induces spindle-like or semi-spindle-like winding patterns. This morphological feature constitutes another significant deviation from the spherically symmetric case. Finally, our analysis of equatorial light deflection in the weak-field limit demonstrates that increasing $Q$ suppresses the deflection angle. Extending this framework to investigate null geodesics in the spacetime of a rotating, charged Kerr--Newman black hole pierced by a cosmic string presents a promising avenue for future research.

	\section*{Appendix A}
	\hspace{1.5em}We discuss the distribution of the roots of the quartic polynomial in Eq. (3.14).
	
	(1) To begin with, we first focus on the case where $\Delta = 0$, which can be divided into two subcases for discussion. (a) For $0 < \varepsilon < 1$, it can be verified that $p, q < 0$, and
	\begin{equation}
		p^2 - 4s = -\frac{2}{\varepsilon^4} \left[ 3 - 4\varepsilon + (8\varepsilon - 9)\sqrt{9 - 8\varepsilon} \right] > 0.\tag{A.1}
	\end{equation}
	According to the theory of root distribution for quartic polynomials, this corresponds to four real roots, two of which are degenerate. (b) For $\varepsilon = 1$, we have $q = s = 0$, and $p < 0$, which also corresponds to four real roots. In this case, the two degenerate real roots are both zero, while the other two distinct real roots are $\pm\sqrt{2}$. In fact, through complex calculations, it can be verified that the four real roots are
	\begin{equation}
		y_{2,3} = \frac{1 - \sqrt{9 - 8\varepsilon}}{2\varepsilon}, \quad
		y_1 = -y_{2,3} - \sqrt{-p - 2y_{2,3}^2}, \quad
		y_4 = -y_{2,3} + \sqrt{-p - 2y_{2,3}^2},\tag{A.2}
	\end{equation}
	which satisfies our aforementioned conclusion, and the distribution of the roots satisfies: $y_1 < -1/\varepsilon < y_{2,3} < y_4$. To summarize, the $\Delta = 0$ case invariably results in four real roots for the quartic polynomial, with two of them degenerate.
	
	(2) For the $\Delta > 0$ case, it can be verified that for $0 < \varepsilon \le 1$, $p < 0$ and $p^2 - 4s > 0$ hold. In this case, the quartic polynomial $P(y)$ possesses four distinct real roots. Numerical verification confirms that the distribution of the four real roots satisfies: $y_1 < -1/\varepsilon < y_2 < y_3 < y_4$.
	
	(3) When $\Delta < 0$, the quartic polynomial possesses two distinct real roots and two complex conjugate roots. In this case, let the two real roots be $y_1$ and $y_2$, which satisfy: $y_1 < -1/\varepsilon < 0 < y_2$.
	
	\section*{Appendix B}
	
	\hspace{1.5em}This appendix presents a summary of the integration results for both $\Delta > 0$ and $\Delta < 0$ cases. To facilitate comparison, the results are formulated using the inverse Weierstrass elliptic function and the elliptic integral of the first kind, with the corresponding parameters explicitly defined. The definitions of $t(x, x_{1(3)})$ appearing in the table are given as follows
	\begin{equation}
		t(x, x_1) = \frac{1}{24}P''(x_1) + \frac{P'(x_1)}{4(x-x_1)},\quad t(x, x_3) = \frac{1}{24}P''(x_3) + \frac{P'(x_3)}{4(x-x_3)}.\tag{B.1}
	\end{equation}
	
	\begin{sidewaystable}[htbp]
		\centering
		\caption*{Table B: Comparison of integral results expressed in two different forms.} 
		\normalsize 
		\renewcommand{\arraystretch}{2.2}      
		\setlength{\extrarowheight}{3pt}       
		\setlength{\arrayrulewidth}{0.3pt}
		\begin{tabular}{lccc}
			\toprule
			& $\Delta < 0$ & \multicolumn{2}{c}{$\Delta > 0$} \\
			\hline
			\addlinespace 
			Distribution of real roots of $P(x)=0$
			& $x_1 < -\frac{1}{3} < x_2$
			& $x_1 < -\frac{1}{3} < x_2 < x_3 < x_4$
			& $x_1 < -\frac{1}{3} < x_2 < x_3 < x_4$ \\
			\hline
			Integration interval
			& $x_\text{init}, x_\text{f} \in \left[-\frac{1}{3}, x_2\right]$
			& $x_\text{init}, x_\text{f} \in \left[-\frac{1}{3}, x_2\right]$
			& $x_\text{init}, x_\text{f} \in \left[x_3, x_4\right]$ \\
			\addlinespace 
			Integral result (in terms of $\wp^{-1}$)
			& $\wp^{-1} \left(t(x, x_1); g_2, g_3 \right) \Big|_{x_\text{f}}^{x_\text{init}}$
			& $\wp^{-1} \left(t(x, x_1); g_2, g_3 \right) \Big|_{x_\text{f}}^{x_\text{init}}$
			& $\wp^{-1} \left(t(x, x_3); g_2, g_3 \right) \Big|_{x_\text{f}}^{x_\text{init}}$ \\
			\addlinespace 
			Distribution of roots of $4t^3 - g_2 t - g_3 = 0$
			& $e_1, e_2 = \overline{e_3} = \alpha + i\beta$
			& $e_1 > e_2 > e_3$
			& $e_1 > e_2 > e_3$ \\
			\addlinespace 
			Definition of parameters
			& \begin{tabular}[c]{@{}l@{}} $g = \frac{1}{2\sqrt{A}},\quad k^2 = \frac{A + \alpha - e_1}{2A}$ \\ $A = \sqrt{(\alpha - e_1)^2 + \beta^2}$ \end{tabular}
			& $k = \sqrt{\frac{e_2 - e_3}{e_1 - e_3}},\quad g = \frac{1}{\sqrt{e_1 - e_3}}$
			& $k = \sqrt{\frac{e_1 - e_3}{e_1 - e_2}},\quad g = \frac{1}{\sqrt{e_1 - e_2}}$ \\
			\addlinespace 
			Integral result (in terms of $F(\varphi, k)$)
			& $g \left[ F(\varphi_2, k) - F(\varphi_1, k) \right]$
			& $g \left[ F(\varphi_1, k) - F(\varphi_2, k) \right]$
			& $g \left[ F(\varphi_1, k) - F(\varphi_2, k) \right]$ \\
			\addlinespace 
			Definition of $\varphi_i$
			& \begin{tabular}[c]{@{}l@{}} $\varphi_1 = \cos^{-1} \sqrt{\frac{t(x_{\text{init}}) - e_1 - A}{t(x_{\text{init}}) - e_1 + A}}$ \\ $\varphi_2 = \cos^{-1} \sqrt{\frac{t(x_{\text{f}}) - e_1 - A}{t(x_{\text{f}}) - e_1 + A}}$ \end{tabular}
			& \begin{tabular}[c]{@{}l@{}} $\varphi_1 = \sin^{-1} \sqrt{\frac{e_1 - e_3}{t(x_{\text{init}}) - e_3}}$ \\ $\varphi_2 = \sin^{-1} \sqrt{\frac{e_1 - e_3}{t(x_{\text{f}}) - e_3}}$ \end{tabular}
			& \begin{tabular}[c]{@{}l@{}} $\varphi_1 = \sin^{-1} \sqrt{\frac{t(x_{\text{init}}) - e_3}{e_2 - e_3}}$ \\ $\varphi_2 = \sin^{-1} \sqrt{\frac{t(x_{\text{f}}) - e_3}{e_2 - e_3}}$ \end{tabular} \\
			\bottomrule
		\end{tabular}
		\label{tab:simple}
	\end{sidewaystable}

	\clearpage 
	


\begin{thebibliography}{99}
		
		\bibitem{1} T. W. B. Kibble, J. Phys. A: Math. Gen. \textbf{9}, 1387 (1976).
		
		\bibitem{2} Juan Deng, Int J Theor Phys \textbf{51}, 1632 (2012).
		
		\bibitem{3} D. Harari and P. Sikivie, Phys. Rev. D \textbf{37}(12), 3438 (1988).
		
		\bibitem{4} A. Vilenkin, Phys. Rev. Lett. \textbf{46}, 1169 (1981).
		
		\bibitem{5} T. W. B. Kibble and N. Turok, Phys. Rev. Lett. \textbf{116B}, 141 (1982).
		
		\bibitem{6} T. Damour and A. Vilenkin, Phys. Rev. Lett. \textbf{85}, 3761 (2000).
		
		\bibitem{7} T. Damour and A. Vilenkin, Phys. Rev. D \textbf{71}, 063510 (2005).
		
		\bibitem{8} Sameer Ahmed, Michael J. Kavic, Steven L. Liebling, Matthew Lippert, Mohammad Mian, John Simonetti, arXiv:2407.04743.
		
		\bibitem{9} J. Ellis, M. Lewicki, C. Lin and V. Vaskonen, Phys. Rev. D \textbf{108}, 103511 (2023).
		
		\bibitem{10} J. J. Blanco-Pillado, K. D. Olum and X. Siemens, Phys. Lett. B \textbf{778}, 392 (2018).
		
		\bibitem{11} D. F. Chernoff, A. Goobar and J. J. Renk, Mon. Not. Roy. Astron. Soc. \textbf{491}, 596 (2020).
		
		\bibitem{12} A. Vilenkin, Y. Levin and A. Gruzinov, JCAP \textbf{11}, 008 (2018).
		
		\bibitem{13} Mukunda Aryal, L. H. Ford, and Alexander Vilenkin, Phys. Rev. D \textbf{34}, 2263 (1986).
		
		\bibitem{14} Ceren H. Bayraktar, Eur. Phys. J. Plus \textbf{133}, 377 (2018).
		
		\bibitem{15} Riasat Ali, Rimsha Babar, Muhammad Asgher, and Tie-Cheng Xia, Int. J. Mod. Phys. A \textbf{37}(17), 2250108 (2022).
		
		\bibitem{16} Wei Zhang, Phys. Lett. B \textbf{861}, 139237 (2025).
		
		\bibitem{17} A. Aliev and D. Gal'tsov, Pis'ma Astron. Zh. \textbf{14}, 116 (1988).
		
		\bibitem{18} D. Gal'tsov and E. Masar, Class. Quantum Grav. \textbf{6}, 1313 (1989).
		
		\bibitem{19} S. Chakraborty and L. Biswas, Class. Quantum Grav. \textbf{13}, 2153 (1996).
		
		\bibitem{20} Eva Hackmann, Betti Hartmann, Claus Lämmerzahl, and Parinya Sirimachan, Phys. Rev. D \textbf{81}, 064016 (2010).
		
		\bibitem{21} Eva Hackmann, Betti Hartmann, Claus Lämmerzahl, and Parinya Sirimachan, Phys. Rev. D \textbf{82}, 044024 (2010).
		
		\bibitem{22} Ishan Swamy, Deobrat Singh, arXiv:2512.08368 [gr-qc].
		
		\bibitem{23} Shao-Wen Wei and Yu-Xiao Liu, Phys. Rev. D \textbf{85}, 064044 (2012).
		
		\bibitem{24} Jingyun Man, Huawen Wang, Hongbo Cheng, arXiv:1010.2308 [gr-qc].
		
		\bibitem{25} Shunichiro Kinoshita, Takahisa Igata, and Kentaro Tanabe, Phys. Rev. D \textbf{94}, 124039 (2016).
		
		\bibitem{26} J. C. Aurrekoetxea, C. Hoy, M. Hannam, Phys. Rev. Lett. \textbf{132}(18), 181401 (2024).
		
		\bibitem{27} E. Hackmann and C. Lämmerzahl, AIP Conf. Proc. \textbf{1577}, 78 (2014).
		
		\bibitem{28} F. Gackstatter, Ann. Phys. \textbf{495}, 352 (1983).
		
		\bibitem{29} Sini R, V. C. Kuriakose, Mod. Phys. Lett. A \textbf{24}(25), 2025 (2009).
		
		\bibitem{30} P. F. Byrd, M. D. Friedman, Handbook of Elliptic Integrals for Engineers and Scientists, Springer, 1971.
		
		\bibitem{31} X. Pang and J. Jia,  Class. Quantum Grav. \textbf{36}, 065012 (2019).
		
	\end{thebibliography}
\end{document}